\documentclass[compact]{jkpapershort}
\usepackage{tikz}
\usetikzlibrary{arrows.meta}
\usetikzlibrary{decorations.pathreplacing}
\usepackage{tikz-cd}
\usepackage{bigints}
\usepackage{biblatex}
\usepackage{aas_macros}
\usepackage{soul}

\makeatletter
\newcommand\footnoteref[1]{\protected@xdef\@thefnmark{\ref{#1}}\@footnotemark}
\makeatother

\newauthornote{\jjvk}{JJVK}{blue!80}

\newcommand{\OIST}{\raisebox{-0.08em}{\includegraphics[height=0.8em]{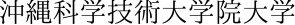}}}

\title{Gravitational entropy is observer\texorpdfstring{\textcolor{black!40!white}{-}}{-}dependent}
\author{Julian De Vuyst\texorpdfstring{\textsuperscript{1,a}}{}, Stefan Eccles\texorpdfstring{\textsuperscript{1,b}}{}, Philipp A.\ H\"ohn,\texorpdfstring{\textsuperscript{1,c}}{} and Josh Kirklin\texorpdfstring{\textsuperscript{1,2,d}}{}}
\institution{\texorpdfstring{\textsuperscript{1}}{}Qubits and Spacetime Unit,\texorpdfstring{\\}{ }Okinawa Institute of Science and Technology \emph{(}\OIST\emph{)},\texorpdfstring{\\}{ }1919-1 Tancha, Onna-son, Kunigami-gun, Okinawa, Japan 904-0495
\texorpdfstring{\\\vspace*{.5em}}{; }\texorpdfstring{\textsuperscript{2}}{}Perimeter Institute for Theoretical Physics,\texorpdfstring{\\}{ }31 Caroline Street North, Waterloo, ON, N2L 2Y5, Canada}

\email{
\textsuperscript{a}\emaillink{julian.devuyst@oist.jp}\hspace*{0.5em}
\textsuperscript{b}\emaillink{stefan.eccles@oist.jp}\hspace*{0.5em}
\textsuperscript{c}\emaillink{philipp.hoehn@oist.jp}\hspace*{0.5em}
\textsuperscript{d}\emaillink{jkirklin@pitp.ca}
}

\abstr{
    In quantum gravity, it has been argued that a proper accounting of the role played by an observer promotes the von Neumann algebra of observables in a given spacetime subregion from Type III to Type II. While this allows for a mathematically precise definition of its entropy, we show that this procedure depends on which observer is employed. We make this precise by considering a setup in which many possible observers are present; by generalising previous approaches, we derive density operators for the subregion relative to different observers (and relative to arbitrary collections of observers and for arbitrary global physical states), and we compute the associated entropies in a semiclassical regime, as well as in some specific examples that go beyond this regime. We find that the entropies seen by distinct observers can drastically differ. Our work makes extensive use of the formalism of quantum reference frames (QRF); indeed, as we point out, the `observers' considered here and in the previous works are nothing but QRFs. In the process, we demonstrate that the description of physical states and observables invoked by Chandrasekaran et al.~\cite{Chandrasekaran:2022cip} is equivalent to the Page-Wootters formalism, leading to the informal slogan ``$\textrm{PW}=\textrm{CLPW}$''. It is our hope that this paper will help motivate a long overdue union between the QRF and quantum gravity communities. Further details will appear in a companion paper. 
}

\bibliography{refs}
\begin{document}
\maketitleandtoc

\section{Introduction}

The generalised entropy formula has long been a striking indication of the deep connection between geometry and information in quantum gravity~\cite{Bekenstein:1972tm,Hawking:1975vcx,Ryu_2006,Lewkowycz_2013,Faulkner_2013,Engelhardt_2015}. Despite being a semiclassical formula, a rigorous understanding of its origin has mostly only been possible (and then even only partially) in certain microscopic theories (for example~\cite{Strominger_1996,Meissner_2004}). This changed with recent work by Chandrasekaran, Longo, Penington, and Witten (CLPW)~\cite{Chandrasekaran:2022cip} (inspired by~\cite{Witten:2021unn,Chandrasekaran_2023,leutheusser2023causal,leutheusser2023emergent}), as well as further investigations by many others (for example~\cite{Jensen:2023yxy,Kudler-Flam:2023qfl,Witten:2023xze, Kudler-Flam:2023hkl,Gesteau:2023hbq,Balasubramanian:2023dpj,Soni:2023fke,AliAhmad:2023etg,Aguilar-Gutierrez:2023odp,Gomez:2023wrq,Gomez:2023upk,Klinger:2023tgi,Klinger:2023auu}), which showed that one may make progress semiclassically by properly accounting for the role of an observer. In particular, the von Neumann algebra of gravitational observables dressed to an observer admits a mathematically well-defined notion of entanglement entropy, and one can in certain cases recover a version of the generalised entropy formula.

What CLPW call an `observer' has been much studied elsewhere in the literature, where it has been called a \emph{quantum reference frame} (QRF) (see~\cite{Aharonov:1967zza,Aharonov:1984,angeloPhysicsQuantumReference2011a,Giacomini:2017zju,delaHamette:2021oex,Hoehn:2023ehz,Hoehn:2019fsy,Hoehn:2020epv,AliAhmad:2021adn,Hoehn:2021flk,delaHamette:2021piz,Vanrietvelde:2018pgb,Vanrietvelde:2018dit,Giacomini:2021gei,Castro-Ruiz:2019nnl,Suleymanov:2023wio,Krumm:2020fws,delaHamette:2020dyi,Kabel:2024lzr,Kabel:2023jve,Carette:2023wpz,Loveridge:2019phw,loveridgeSymmetryReferenceFrames2018a,Bartlett:2006tzx,Castro-Ruiz:2021vnq}, among many others). The QRF and quantum gravity research programs have largely been carried out in parallel, without as much overlap as one might expect, given the key role played by reference frames in the foundations of gravity. It is therefore rather exciting that there is now an opportunity for a long overdue union between these two communities (and indeed this union has already begun to bear fruit~\cite{Goeller:2022rsx,Carrozza:2022xut,Carrozza:2021gju,Kabel:2024lzr,Kabel:2023jve,Susskind:2023rxm,Fewster:2024pur,Gomez:2023upk,KirklinGSL}).

Although they did not explicitly state it, much of the construction of CLPW is carried out \emph{within the perspective} of a reference frame, via a so-called Page-Wootters {(PW)} reduction~\cite{Page:1983uc,1984IJTP...23..701W}. This reduction is a key part of the \emph{perspective-neutral} approach to QRFs \cite{delaHamette:2021oex,Hoehn:2023ehz,Hoehn:2019fsy,Hoehn:2020epv,AliAhmad:2021adn,Hoehn:2021flk,Giacomini:2021gei,Castro-Ruiz:2019nnl,delaHamette:2021piz,Vanrietvelde:2018dit,Vanrietvelde:2018pgb,Hoehn:2021flk,Suleymanov:2023wio}. Thus, the observation that~\cite{Chandrasekaran:2022cip} was secretly a paper about quantum reference frames may be summarised by the equation\footnote{Of course,~\cite{Chandrasekaran:2022cip} makes many non-trivial additional observations, so this equation should not be interpreted too literally.}
\begin{equation}
    \text{PW} = \text{CLPW}.
    \label{Equation: PW=CLPW}
\end{equation}
We will explain this connection in more detail later in the paper.

The version of entropy formulated by CLPW is clearly observer-dependent. In this note, we will explain how this may be understood using features of the QRF formalism. A QRF is a set of degrees of freedom playing the role of the observer; there are many QRFs in the universe, and hence many different entropies. We will describe how these entropies can drastically differ, in a way that depends on the precise nature of the QRF one uses. The observer-dependence of entropy may be alternately stated as a \emph{relativity} of entropy. This is extremely natural in gravity, where it seems entropy, like almost everything else, must depend on the frame of reference. 
Technically, it is a consequence of \emph{subsystem relativity}: the observation that different QRFs decompose a composite system subject to gauge symmetries in different ways into gauge-invariant subsystems \cite{Hoehn:2023ehz,AliAhmad:2021adn,delaHamette:2021oex,Castro-Ruiz:2021vnq}. 

We will generalise the framework of~\cite{Chandrasekaran:2022cip,Jensen:2023yxy} to the case where there are many QRFs in the universe. We will find the density operators for a subregion relative to any arbitrary subset of these QRFs. Our formula for the density operators holds for arbitrary states, in contrast with previous work, and furthermore it is exact. We will also compute the corresponding von Neumann entropies of these density operators in a generalisation of the semiclassical regime described by~\cite{Chandrasekaran:2022cip,Jensen:2023yxy}, and some leading order corrections to this regime in a specific example. In addition, we will consider what we call an \emph{anti}semiclassical regime, in which a given QRF does not allow one to observe much about the state of the subregion. Finally, we will discuss how different regimes can apply to different QRFs within the \emph{same} state, thus leading to an explicit demonstration of the observer-dependence of gravitational entropy.

To be clear, the total entropy of \emph{all} the degrees of freedom in a subregion, including every QRF in that subregion, is \emph{not} observer-dependent.\footnote{For subregions with horizons this is a natural definition of horizon entropy, which is therefore observer-independent.} The observer-dependence comes in when we \emph{restrict} which QRFs one can use to measure the fields in the subregion. The choice of QRFs one restricts to is a purely operational input in our approach.

Another important point to make is that, although we consider a system with many QRFs, we will always assume that each QRF has access to the fields within a single fixed subregion. More generally, different QRFs will have access to different subregions; this will clearly be a source of observer-dependence in the entropy, since the state of the fields in one subregion can be completely different from that in another. However, such an observer-dependence of entropy is not one special to internal QRFs but already arises for classical external frames whose dynamics is not included in the description. The paradigmatic example of this concerns the different descriptions of quantum states by inertial and Rindler observers owing to the \emph{differing} regions and thus degrees of freedom to which they each have access. This leads to a dependence of entropy on such non-dynamical (in this sense external) observers that can be extended to general spacetimes \cite{Marolf:2003sq,Ju:2023bjl,Ju:2023dzo}. On the other hand, the observer-dependence in the entropy that we study in this paper is qualitatively very different and refers to the quantum field degrees of freedom within the \emph{same} subregion. It is a consequence of QRFs being \emph{dynamical} frames and will have to do with their intrinsic properties, and correlations with the other degrees of freedom present in the system.

We hope that this relatively informal paper will provide a comfortable meeting point for the quantum gravity and quantum reference frame communities. A more detailed account with many other results and examples may be found in the longer companion work~\cite{DVEHK}. Similarly, an in-depth discussion of how the observer/QRF setup should be understood from the perspective of gravitational perturbation theory -- and specifically of its relation to linearisation instabilities -- can be found in \cite{DeVuyst:2024grw}.

The present paper proceeds as follows. In Section~\ref{Section: the system}, we will construct the effective description of the low energy gravitational system we study in this paper: a quantum field theory coupled to some number of observers carrying clocks. We will describe relevant details of the perspective-neutral approach to QRFs, and explain in greater detail the significance of Eq.~\eqref{Equation: PW=CLPW}. Then, in Section~\ref{Section: entropy}, we will explain how to compute the density operator relevant when we use some subset of the clocks to observe the QFT in a subregion, and we describe and compute the entropy in the semiclassical and antisemiclassical regimes. We use these results to properly demonstrate the observer-dependence of gravitational entropy in Section~\ref{Section: subsystem relativity}, before finally concluding in Section~\ref{Section: conclusion}.

\noindent{\textbf{Note added:} While completing this work and its detailed version \cite{DVEHK}, Ref.~\cite{Fewster:2024pur} appeared, which similarly points out that observers in CLPW \cite{Chandrasekaran:2022cip} are QRFs. The key difference between our work and \cite{Fewster:2024pur} is that the authors of the latter invoke the so-called operational approach to QRFs \cite{Carette:2023wpz,loveridgeSymmetryReferenceFrames2018a}, which constructs gauge-invariant algebras but does not implement constraints on states. However, in quantum gravity, constraints are imposed on states as well, and going this extra step means invoking the perspective-neutral approach to QRFs and is what allows us to establish `equation'~\eqref{Equation: PW=CLPW}. Another difference is that we study multiple observers, traces and entropies. Notwithstanding, at the kinematical algebraic level,  \cite{Fewster:2024pur} is compatible with our approach and we look forward to comparing our works. }

\section{Low energy gravitational system: fields and clocks}
\label{Section: the system}

We begin by applying the formalism of quantum reference frames (QRFs) to perturbative gravity in a local quantum field theory setting. Unlike typical QRF settings, this will be a hybrid scenario, where the frames are quantum mechanical and the ``observed system'' is a set of quantum fields containing gravitons.\footnote{As indicated in the Introduction, we employ the so-called \emph{perspective-neutral} approach to QRFs \cite{delaHamette:2021oex,Hoehn:2023ehz,Hoehn:2019fsy,Hoehn:2020epv,periodic,AliAhmad:2021adn,Hoehn:2021flk,Giacomini:2021gei,Castro-Ruiz:2019nnl,delaHamette:2021piz,Vanrietvelde:2018dit,Vanrietvelde:2018pgb,Hoehn:2021flk,Suleymanov:2023wio}, see \cite[Sec.~II]{Hoehn:2023ehz} for a gentle introduction and a comparison with special covariance. A few other QRF formalisms have emerged, including the purely perspectival one \cite{Giacomini:2017zju,delaHamette:2020dyi,Kabel:2024lzr,Krumm:2020fws}, the operational one \cite{Carette:2023wpz,Loveridge:2019phw,loveridgeSymmetryReferenceFrames2018a} and the quantum information one \cite{Castro-Ruiz:2021vnq,Bartlett:2006tzx}. Differences are rooted in how the approaches treat symmetries. The perspective-neutral approach is singled out by treating `external' frame transformations as gauge, which means  it implements constraints as in gauge theory and gravity. It is thus the one applicable to the present gravitational scenario. (For the special case of ideal QRFs, the purely perspectival approach is equivalent to the one used here, not however for general QRFs \cite{delaHamette:2021oex}.) } More precisely, let us consider a low energy gravitational system consisting of quantum reference frames $C_i$, $i=1,\dots ,n$, for some arbitrary $n\in\NN$, they aren't coupled at the order we consider and an effective quantum field theory in the $G_N\to0$ limit. The kinematical Hilbert space of this system decomposes as
\begin{equation}\label{Hkin}
    \mathcal{H}_{\text{kin}} = \mathcal{H}_S \otimes \bigotimes_{i=1}^n \mathcal{H}_{i},
\end{equation}
where $\mathcal{H}_S$ is the kinematical Hilbert space of the QFT, and  $\mathcal{H}_{i}$ is the kinematical Hilbert space of $C_i$. 

Each QRF $C_i$ will carry a clock whose evolution is generated by its Hamiltonian $H_i$. For simplicity, this is all the QRF will carry, and furthermore, the spectrum of each $H_i$ will be taken to be non-degenerate and continuous. We can then write $\mathcal{H}_{i} = L^2(\sigma_i)$, where $\sigma_i\subset\RR$ is the energy spectrum of $C_i$. {We briefly comment on the degenerate case later and discuss it in more detail in \cite{DVEHK}. In our exposition, we will use the clock QRF formalism from \cite{Hoehn:2019fsy,Hoehn:2020epv}.\footnote{The present QRFs are ones associated with the translation group. For the generalisation of the below perspective-neutral QRF formalism to general unimodular groups see \cite{delaHamette:2021oex} and specifically for periodic clocks see \cite{periodic}.}} One may construct a set of states
\begin{equation}\label{clockstate}
    \ket{t}_i = \frac1{\sqrt{2\pi}}\int_{\sigma_i}   e^{-i\varepsilon t}\ket{\varepsilon}_i, \qquad t\in\RR,
\end{equation}
for each frame, where $\ket{\varepsilon}_i$, $\varepsilon\in\sigma_i$ are energy eigenstates.\footnote{{The states $\ket{t}_i$ are not normalisable, and so in fact are not truly elements of $\mathcal{H}_i$, unless $\sigma_i$ is bounded in both directions. Moreover, when $\sigma_i$ is bounded, the $\ket{t}_i$ are neither perfectly distinguishable, nor eigenstates of a clock operator. Nevertheless, the clock states give rise to well-defined clock observables using the formalisms of covariant POVMs \cite{Hoehn:2019fsy,Hoehn:2020epv,Loveridge:2019phw,Smith2019quantizingtime,holevoProbabilisticStatisticalAspects1982,Busch1994,Braunstein:1995jb}(see also \cite{DVEHK} for the present case). In general, one may also include an additional arbitrary phase factor $e^{ig(\varepsilon)}$ in the integrand defining the states $\ket{t}_i$ \cite{Hoehn:2019fsy,Braunstein:1995jb}. We choose here to absorb this phase factor into the energy eigenstates.}} These have the interpretation that, in the state $\ket{t}_i$, the time given by clock $C_i$ is $t$. The states $\ket{t}_i$ furnish a resolution of the identity: 
\begin{equation}
    \mathds{1}_{i} = \int_{\mathbb{R}}\dd{t}\dyad{t}_i
\end{equation}
(where $\dyad{t}_i$ is shorthand for $\ket{t}_i\!\bra{t}_i$), and they transform covariantly:
\begin{equation}\label{cstatecov}
    e^{-iH_i\tau} \ket{t}_i = \ket{t+\tau}_i.
\end{equation}
These states are typically not orthogonal for different clock readings, i.e.\ for $t\ne t'$, we typically get $\braket{t}{t'}\ne0$. In this sense, the time given by the clock is not definite but rather has some fuzziness. Quantum reference frames with this property are called `non-ideal'; conversely, quantum reference frames with $\braket{t}{t'}=0$ if $t\ne t'$ are called `ideal'. For the simple kind of QRF we are considering, a clock is ideal if its energy spectrum is the whole real line, and non-ideal otherwise.

To get the physical Hilbert space of this system, one must impose the gauge constraints. In gravity, this includes diffeomorphisms. At the level of the low energy gravitational system  described above, diffeomorphisms act not only on the fields but also the clocks, changing the times that they read. Like  \cite{Chandrasekaran:2022cip,Jensen:2023yxy}, we will for simplicity focus on a single constraint and assume that all remaining pertinent constraints have already been imposed with $\mathcal{H}_{\text{kin}}$ the resulting invariant Hilbert space (more on this shortly):
\begin{equation}\label{constraint1}
    H = H_S + \sum_{i=1}^n H_i,
\end{equation}
where $H_S$, an operator acting on $\mathcal{H}_S$, is the generator of the action of a certain diffeomorphism on the effective field degrees of freedom.\footnote{Here we are ignoring any explicit interaction terms between the QFT and the clocks, {in line with a $G_N\to0$ limit}.} Overall, $H$ may be thought of as the generator of this diffeomorphism on the full system, accounting for its action on the QRFs as well as the fields. We impose invariance under this diffeomorphism by setting $H=0$. 

\begin{figure}
    \centering
    \begin{tikzpicture}
        \fill[blue!10] (3,0) -- (6,3) -- (9,0) -- (6,-3) -- cycle;
        \fill[red!10] (3,0) -- (0,3) -- (0,-3) -- cycle;
        \fill[red!10] (9,0) -- (12,3) -- (12,-3) -- cycle;
    
        \draw (0,-3) -- (6,3) -- (12,-3);
        \draw (0,3) -- (6,-3) -- (12,3);

        \begin{scope}[blue, thick]
            \foreach \i in {3,4,5,6,7} {
                \draw (6,3) .. controls ({1+\i},0.5) .. ({1+\i},0);
                \draw[-{Stealth[scale=1.1,angle'=45]}] (6,-3) .. controls ({1+\i},-0.5) .. ({1+\i},0.1) node[left] {\footnotesize$C_\i$};
            }
        \end{scope}
        \begin{scope}[red, thick]
            \foreach \i in {8,9} {
                \draw (12,-3) .. controls ({2+\i},-0.5) .. ({2+\i},0);
                \draw[-{Stealth[scale=1.1,angle'=45]}] (12,3) .. controls ({2+\i},0.5) .. ({2+\i},-0.1) node[left] {\footnotesize$C_\i$};
            }
            \foreach \i in {1,2} {
                \draw (0,-3) .. controls ({\i},-0.5) .. ({\i},0);
                \draw[-{Stealth[scale=1.1,angle'=45]}] (0,3) .. controls ({\i},0.5) .. ({\i},-0.1) node[left] {\footnotesize$C_\i$};
            }
        \end{scope}

        \node[above right, blue] at (7.5,1.5) {\Large$\mathcal{U}$}; 
        \node[above right, red] at (1.5,1.5) {\Large$\mathcal{U}'$}; 
        \node[above left, red] at (10.5,1.5) {\Large$\mathcal{U}'$}; 
    \end{tikzpicture}
    \caption{We consider a low energy gravitational system made up of field degrees of freedom described by an effective QFT, and an arbitrary number of clock degrees of freedom, each of which is carried by a quantum reference frame $C_i$. The clocks are taken to measure time along the modular flow of some fixed QFT state $\ket{\psi_S}$ with respect to the algebra $\mathcal{A}_{\mathcal{U}}$ of QFT operators in a fixed spacetime region $\mathcal{U}$. Heuristically (but not always literally), one may imagine that the clocks evolve along some worldlines resembling the integral curves of a boost in $\mathcal{U}$ or its complement $\mathcal{U}'$. By the timelike tube theorem \cite{Borchers1961,Araki1963AGO,Witten:2023qsv,Strohmaier:2023opz}, each QRF in $\mathcal{U}$ has access to the full regional QFT algebra $\mathcal{A}_\mathcal{U}$.}
    \label{Figure: observers in subregion}
\end{figure}
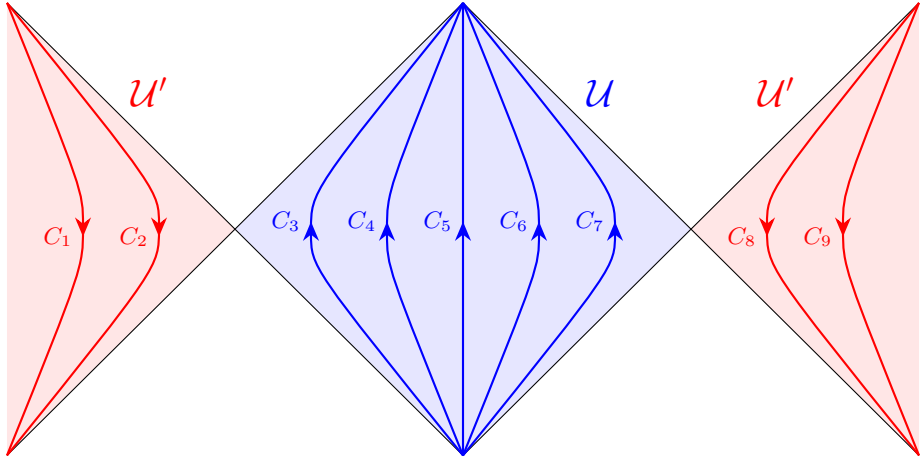

We will in particular assume that $\beta H_S$ for some positive $\beta\in\RR$ is the modular Hamiltonian of some fixed cyclic and separating QFT state $\ket{\psi_S}\in\mathcal{H}_S$ {(thus a KMS state \cite{haag2012local,WittenRevModPhys.90.045003}), for} the von Neumann algebra of bounded QFT operators $\mathcal{A}_{\mathcal{U}}\subset \mathcal{B}(\mathcal{H}_S)$ with support in some spacetime region $\mathcal{U}$. Hence, imposing the constraint $H=0$ means that each of the clocks $C_i$ measures \emph{modular time} with respect to $\ket{\psi_S}$ and $\mathcal{A}_{\mathcal{U}}$. One may heuristically think of the QRFs as evolving along worldlines generated by the modular flow of the QFT (see Fig.~\ref{Figure: observers in subregion}). We can imagine that each of the clocks located within $\mathcal{U}$ is capable of measuring the fields in the vicinity of its worldline, which implies they can measure everything in $\mathcal{A}_{\mathcal{U}}$ by the timelike tube theorem~\cite{Borchers1961,Araki1963AGO,Witten:2023qsv,Strohmaier:2023opz}. This motivation does not apply to clocks located outside of $\mathcal{U}$ -- but there is no reason we cannot also use those clocks in what follows.

The simplest version of this setup, where modular flow admits a geometric interpretation as a diffeomorphism, is where $\mathcal{U}$ is a Rindler wedge in Minkowski spacetime, and $\ket{\psi_S}$ is just the QFT vacuum state~\cite{Bisognano:1975ih} (this is one manifestation of the Unruh effect~\cite{UnruhEffect}). This observation has been generalised to general Killing horizons with $H_S$ the associated boost Hamiltonian \cite{Sewell:1982zz}, and a conjecture of~\cite{Jensen:2023yxy}, based on evidence such as~\cite{Sewell:1982zz,Bisognano:1975ih,Hislop:1981uh,Borchers:1998ye,Casini_2011,Wong:2013gua,Cardy:2016fqc}, posits that a $\ket{\psi_S}$ with the right properties always exists in the general case (this has been further explored in~\cite{sorce2024analyticity}). Our setup is thus safe for regions bounded by Killing horizons, in which case the clocks measure the Killing boost time, and may apply more generally, pending a clarification of the status of the geometric modular flow conjecture in \cite{Jensen:2023yxy}. Note that here $\ket{\psi_S}$ is a  state for the fields in the \emph{interior} of spacetime, but not necessarily on its asymptotic boundary. Boundary degrees of freedom can be modeled by the inclusion of an additional (ADM) clock, as in~\cite{Chandrasekaran:2022cip,Jensen:2023yxy,Kudler-Flam:2023qfl}.

As pointed out in \cite{Jensen:2023yxy,Kudler-Flam:2023qfl,DeVuyst:2024grw}, the constraint in Eq.~\eqref{constraint1} is of \emph{second order} in $\sqrt{G_N}$ in the gravitational perturbation theory around a background, in line with being a Hamiltonian including the gravitons. A clear justification for imposing it nevertheless in the $G_N\to0$ limit (where otherwise the linear order would suffice) comes from the theory of linearisation instabilities when $H$ is an isometry generator in a spatially closed universe such as the de Sitter space in \cite{Chandrasekaran:2022cip}, where it turns out to be a necessary stability condition ensuring perturbative consistency \cite{DeVuyst:2024grw}. In this case, the remaining pertinent constraints that are assumed to already be imposed on $\mathcal{H}_{\rm kin}$ and $\mathcal{A}_\mathcal{U}$ correspond to all linearised diffeomorphism and all matter gauge constraints, as well as any additional isometry generator (which necessarily is of second order as well). In all other cases, the imposition of $H$ on the $G_N\to0$ theory stands on somewhat weaker footing and must be justified differently \cite{DeVuyst:2024grw}.

Imposing $H=0$ is not as simple as restricting to the $H=0$ subspace of $\mathcal{H}_{\text{kin}}$, because this subspace typically will contain far fewer states than is physically desirable.\footnote{This is a consequence of the continuous nature of the spectrum of $H$, which causes otherwise physically reasonable gauge-invariant states to not be normalizable.} Instead, one can impose the constraint by introducing a new inner product involving an average over gauge transformations (the `group-averaging' inner product):
\begin{equation}
    \Braket{\phi_1}{\phi_2} = 2\pi\bra{\phi_1}\delta(H)\ket{\phi_2} = \int_{\mathbb{R}}\dd{t}\mel{\phi_1}{e^{-iHt}}{\phi_2},
    \label{Equation: physical inner product}
\end{equation}
where $\ket{\phi_1},\ket{\phi_2}\in\mathcal{H}_{\text{kin.}}$. Since one has a new inner product, one also has a new Hilbert space on which it is defined, and this is the space of physical states $\mathcal{H}_{\text{phys}}$. There are various {equivalent} ways to directly construct $\mathcal{H}_{\text{phys}}$. For example, in refined algebraic quantisation~\cite{Marolf:2000iq, Giulini:1998kf,Giulini:1998rk}, physical states are distributions on the kinematical Hilbert space obtained using a so-called `rigging map'; these distributions solve the constraint $H=0$. In the formalism of coinvariants~\cite{Chandrasekaran:2022cip,AHiguchi_1991}, a physical state is an equivalence class of kinematical states modulo gauge transformations. In any case, there is a linear map $M:\mathcal{H}_{\text{kin}}\to\mathcal{H}_{\text{phys}}$ taking a given kinematical state to the corresponding physical state, and for our purposes {here} this abstract description will suffice{; a more detailed discussion (including the equivalence of the two ways of imposing constraints)} is deferred to \cite{DVEHK}. Two kinematical states related by a gauge transformation map to the same physical state under $M$, and it implements the constraint via $MH=0$. To distinguish physical states from kinematical states, we use the notation $\Ket{\phi}$ for the former, while we use the standard notation $\ket{\phi}$ for the latter, and we will commonly set $\Ket{\phi}=M\ket{\phi}$. In Eq.~\eqref{Equation: physical inner product}, we have $\Ket{\phi_1}=M\ket{\phi_1}$ and $\Ket{\phi_2}=M\ket{\phi_2}$.

\subsection{Observables and perspectives}

The main subject of interest in the present paper comprises the algebra of physical observables of the subregion $\mathcal{U}$. However, there are different ways one can choose to construct this algebra. Let us here describe three: first, the algebra made up purely of fields in the subregion; second, the enlarged algebra one obtains by `dressing' field observables to a clock $C_i$; and last, the further enlarged algebra one obtains by allowing for `reorientations' of that clock.

The simplest way to try to make a physical observable is to take any $O\in\mathcal{A}_{\mathcal{U}}$ which commutes with $H_S$. Although $O$ is defined as a kinematical operator, any such gauge-invariant $O$ has an unambiguous representation $r(O)$ on physical states defined by $r(O)\Ket{\phi} = \Ket{O\phi}=MO\ket{\phi}$, or simply $r(O) M = M O$. Thus, the subalgebra of $\mathcal{A}_{\mathcal{U}}$ consisting of gauge-invariant operators
\begin{equation}
    \mathcal{A}_{\mathcal{U}}^H := \Big\langle O\in\mathcal{A}_\mathcal{U}\mid [O,H_S]=0\Big\rangle\,,
\end{equation}
where $\langle A,B\rangle$ denotes the von Neumann algebra generated by operators of the form $A$ and $B$, has a representation on $\mathcal{H}_{\text{phys}}$, which gives an algebra of physical observables $r(\mathcal{A}_{\mathcal{U}}^H)$ for $\mathcal{U}$.

Unfortunately, the algebra $\mathcal{A}_{\mathcal{U}}^H$ is often hopeless for the purpose of definining a physical subsystem. $\mathcal{A}_{\mathcal{U}}$ is a Type $\mathrm{III}_1$ {factor}\footnote{A factor is an algebra with trivial centre, meaning only multiples of the identity commute with all other elements.} \cite{Araki:1976zv,WittenRevModPhys.90.045003,Yngvason_2005,Jensen:2023yxy}, and a mathematical fact about such algebras is that they always admit ergodic modular flows \cite{Connes,Takesaki1979}, with the \emph{vacuum} modular flow typically one of them,\footnote{We thank Elliot Gesteau for discussion on this.} which means that the only operators they contain which are invariant under such a flow are constant multiples of the identity; indeed, this is the case with the Bunch--Davies vacuum in de Sitter space \cite{Chandrasekaran:2022cip,Chen:2024rpx}. Hence, in that case we have $\mathcal{A}_{\mathcal{U}}^H = \mathbb{C}\mathds{1}$. This algebra does not allow us to observe any physics whatsoever. For other types of KMS states, the modular flow need not necessarily be ergodic, but $\mathcal{A}^H_\mathcal{U}$ may still be `too small' to be physically desirable.

One way out is to use the clocks. In particular, we can take a (not necessarily gauge-invariant) QFT operator $a\in\mathcal{A}_{\mathcal{U}}$ and \emph{dress} it to a given clock $C_i$, which works as follows. First, one considers the operator $a \otimes \dyad{\tau}_i$, which {conditions on} clock $C_i$ such that its time reads $\tau$, while simultaneously acting with $a$ on the QFT. This by itself is not gauge-invariant; one gets a gauge-invariant operator by constructing its `$G$-twirl':
\begin{equation}
    O^\tau_{C_i}(a) = \int_{\mathbb{R}} \dd{t} e^{-i(H_S+H_i)t}\qty(a \otimes \dyad{\tau}_i) e^{i(H_S+H_i)t}.
\end{equation}
Since this operator commutes with the constraint, it yields a well-defined operator\footnote{The difference between the kinematical and physical representation of these operators is essentially the one between incoherent and coherent group averaging, e.g., see \cite{delaHamette:2021oex}.} 
\begin{equation}
    \mathcal{O}_{C_i}^\tau(a) = r(O^\tau_{C_i}(a))
\end{equation}
acting on physical states. {This operator is known both as a `dressed observable', because it is constructed by dressing the fields with the clock, and also as a `relational observable' \cite{Rovelli:1990pi,Marolf:1994ss,Chataignier:2019kof,Hoehn:2019fsy}, because it measures the field observable $a$ when the clock reads $\tau$. Indeed, it was shown in \cite{Hoehn:2019fsy} that these operators are a quantisation of the classical relational observables in \cite{Dittrich:2004cb}.} 

To get a more explicit interpretation of this observable, we can imagine being in the \emph{perspective} of the reference frame $C_i$. In particular, a given physical state $\Ket{\phi} = M \ket{\phi}$ may always be written in the form $\Ket{\phi} = M(\ket{\tau}_i\otimes\ket*{\phi_{|i}(\tau)})$ (up to reordering of kinematical tensor factors), with the state $\ket*{\phi_{|i}(\tau)}$ given by the so-called `Page-Wootters (PW) reduction' of $\Ket{\phi}$:\footnote{{In most formulations of the PW formalism, this is written as a conditioning of physical states on the clock reading $\tau$: $\bra{\tau}\phi)$, e.g.\ see \cite{Page:1983uc,1984IJTP...23..701W,giovannettiQuantumTime2015,Smith2019quantizingtime}. Comparing with Eq.~\eqref{PWred}, this standard form of the PW reduction thus identifies physical states as $\Ket{\phi}=\Pi_{\rm phys}\ket{\phi}$, where $\Pi_{\rm phys}=\int\dd{t}\exp(-iHt)$ is the coherent group averaging operator of refined algebraic quantization, and the conditioning as $\mathcal{R}_i(\tau)=\bra{\tau}_i\otimes\bigotimes_{j\neq i}\mathds{1}_j\otimes \mathds{1}_S$ \cite{Hoehn:2019fsy,Hoehn:2020epv}. (In fact, as physical states are kinematical distributions in this approach, a more accurate description is $\Bra{\phi}=\Pi_{\rm phys}\ket{\psi}$ \cite{Giulini:1998kf}.) Our definition of the PW reduction map in Eq.~\eqref{PWred} is equivalent but agnostic to which method is used for constraint implementation.}}
\begin{equation}\label{PWred}
    \ket*{\phi_{|i}(\tau)} = \mathcal{R}_i(\tau) \Ket{\phi} := \bra{\tau}_i \int_{\mathbb{R}}\dd{t} e^{-iHt}\ket{\phi}\quad \in \quad \mathcal{H}_S \otimes\bigotimes_{\substack{j=1\\j\ne i}}^n\mathcal{H}_{j}.
\end{equation}
 It may be shown \cite{Hoehn:2019fsy,Hoehn:2020epv,delaHamette:2021oex} that the expectation value of $\mathcal{O}^\tau_{C_i}(a)$ in the state $\Ket{\phi}$ is given by
\begin{equation}
    \Mel{\phi}{\mathcal{O}^\tau_{C_i}(a)}{\phi} = \mel*{\phi_{|i}(\tau)}{a}{\phi_{|i}(\tau)}.
\end{equation}
A key facet of the QRF formalism is to think of $\ket*{\phi_{|i}(\tau)}$ as the state of the system \emph{in the perspective of $C_i$} (this is what the notation `$|i$' denotes) when its clock reads $\tau$. Thus, the right hand side measures the expectation value of $a$ \emph{in the perspective of $C_i$}. 

Note that $\ket*{\phi_{|i}(\tau)}$ takes the form of a state for the other QRFs, and the QFT. It can be shown that the so-called \emph{reduction} map 
\begin{equation}
    \mathcal{R}_i(\tau):\mathcal{H}_{\text{phys}}\to \mathcal{H}_S \otimes \bigotimes_{j\ne i}\mathcal{H}_{j}
\end{equation}
is an isometry \cite{Hoehn:2019fsy,Hoehn:2020epv,delaHamette:2021oex} {(in fact, it is a unitary gauge fixing)}. The physical Hilbert space is thus isometric to the image of $\mathcal{R}_i(\tau)$, which is typically a subspace
\begin{equation}\label{perspH}
    \mathcal{H}_{|i} = \Pi_{|i}\qty\bigg(\mathcal{H}_S \otimes \bigotimes_{j\ne i}\mathcal{H}_{j})\subseteq \mathcal{H}_S \otimes \bigotimes_{j\ne i}\mathcal{H}_{j},
\end{equation}
where $\Pi_{|i} = \mathcal{R}_i(\tau)\mathcal{R}_i(\tau)^\dagger$. Explicitly, one may show that
\begin{equation}\label{proj}
    \Pi_{|i} = \int_{\sigma_i} \delta(H_S+\sum_{j\ne i}H_j+\varepsilon) \dd{\varepsilon}.
\end{equation}
In words, $\Pi_{|i}$ is a projection onto the subspace in which $H_S+\sum_{j\ne i}H_j\in -\sigma_i$.\footnote{For an ideal frame, $\Pi_{|i}$ is the identity, but for a non-ideal frame the projection is non-trivial.} This projection $\Pi_{|i}$ is what remains of the constraint $H=0$, and the state $\ket*{\phi_{|i}(\tau)}$ must be in its image $\mathcal{H}_{|i}$, which is known as the `reduced Hilbert space' in the perspective of $C_i$. The original physical Hilbert space constructed above $\mathcal{H}_{\text{phys}}$ is sometimes conversely known as the `perspective-neutral' Hilbert space, because it treats all of the clock QRFs on an equal footing {and links their respective internal perspectives via QRF transformations built from the reduction maps \cite{Hoehn:2019fsy,Hoehn:2020epv,delaHamette:2021oex,Hoehn:2023ehz,DVEHK,Vanrietvelde:2018pgb}}. 

A general QFT operator $a\in\mathcal{A}_{\mathcal{U}}$ acting in the perspective of $C_i$ will not keep us in the physical subspace $\mathcal{H}_{|i}$, so one should restrict to those which do, i.e.\ those satisfying $[a,\Pi_{|i}]=0$.\footnote{Of course, for $[a,\Pi_{|i}]\ne 0$, the dressed observable $\mathcal{O}_{C_i}^\tau(a)$ is still a perfectly legitimate operator at both the perspective-neutral and reduced levels. However, including such observables at this stage would lead to certain issues stemming from the fact that the set of all dressed observables for arbitrary $a$ does not close as an algebra, and so would not define a quantum subsystem in a traditional sense.}
Then it may be shown that \cite{Hoehn:2019fsy,DVEHK}
\begin{equation}
    \mathcal{O}^\tau_{C_i}(a) = \mathcal{R}_i(\tau)^\dagger a \mathcal{R}_i(\tau).
\end{equation}
Thus, the interpretation of $\mathcal{O}^\tau_{C_i}(a)$ extends beyond just expectation values: $\mathcal{O}^\tau_{C_i}(a)$ is exactly the physical operator corresponding to $a$ in the perspective of $C_i$ at time $\tau$. Thus, the relational observable $\mathcal{O}^\tau_{C_i}(a)$ is the \emph{perspective-neutral} implementation of this operator.  Furthermore, the set of all such operators
\begin{equation}
    \mathcal{O}_{\mathcal{U}|i} = \Big\langle\mathcal{O}^\tau_{C_i}(a) \mid a\in\mathcal{A}_{\mathcal{U}}, \,[a,\Pi_{|i}]=0\Big\rangle
\end{equation}
forms an algebra isomorphic via $\mathcal{O}^\tau_{C_i}(a)\leftrightarrow \Pi_{|i}a$ to the algebra $\Pi_{|i}\mathcal{A}_{\mathcal{U}}^{\Pi_{|i}}$ of QFT operators in $\mathcal{U}$ commuting with $\Pi_{|i}$, projected onto $\mathcal{H}_{|i}$. We have not written superscripts ${}^\tau$ on the algebra $\mathcal{O}_{\mathcal{U}|i}$ or the projection $\Pi_{|i}$, because it turns out these do not depend on~$\tau$.

Since we are already involving the frame $C_i$, there is one more important kind of physical operator we can consider: $V_i(t) = r(e^{-iH_it})$, for $t\in \RR$. The operator $e^{-iH_it}$ acts on kinematical states by changing the time of clock $C_i$ via $\ket{\tau}_i\to\ket{\tau+t}_i$, and $V_i(t)$ is the physical version of this change. In the field of QRFs, this is called a `reorientation' of $C_i$ \cite{delaHamette:2021oex,Hoehn:2023ehz}. Involving reorientations allows for a beautiful picture of \emph{relational quantum dynamics}, supported by equations such as the following:
\begin{align}
    \mathcal{R}_i(\tau) V_i(t) \Ket{\phi} &= \ket*{\phi_{|i}(\tau-t)} = e^{i(H_S+\sum_{j\ne i}H_j)t} \ket*{\phi_{|i}(\tau)},\\
    V_i(t) \mathcal{O}_{C_i}^\tau(a) V_i(t)^\dagger &= \mathcal{O}_{C_i}^{\tau-t}(a) = \mathcal{O}_{C_i}^{\tau}\qty(e^{i(H_S+\sum_{j\ne i}H_j)t}ae^{-i(H_S+\sum_{j\ne i}H_j)t}),
\end{align}
which have the interpretation of a Schr\"odinger evolution for the states in the perspective of $C_i$ {(this is the standard PW formalism \cite{Page:1983uc,1984IJTP...23..701W})}, and a Heisenberg evolution for the operators dressed to $C_i$. The Hamiltonian for this evolution is $H_S+\sum_{j \ne i}H_j = H-H_i$, and in the perspective of $C_i$, the reorientation $V_i(t)$ maps to a time evolution in the opposite direction, $\exp(i(H_S+\sum_{j \ne i}H_j)t)$.\footnote{Unlike the more ordinarily studied quantum mechanical case~\cite{Hoehn:2019fsy,Hoehn:2020epv,AliAhmad:2021adn,Giacomini:2021gei,Castro-Ruiz:2019nnl,periodic}, there is no dressed operator which could also generate this time evolution, because the modular flow of Type III algebras is an outer automorphism.} Further details may be found in~\cite{DVEHK}, but for now, suffice it to say that if we want to allow the fields in $\mathcal{U}$ to be dressed to $C_i$ as described above, while allowing the subsystem so described to be dynamical, then we have to account for reorientations.

Thus, we can describe the fields in $\mathcal{U}$ relative to a given clock $C_i$ as a a dynamical subsystem in terms of the algebra of physical observables generated by dressed observables $\mathcal{O}^\tau_{C_i}(a)$ and reorientations $V_i(t)$.\footnote{At this point, we \emph{can} include dressed observables $\mathcal{O}^\tau_{C_i}(a)$ with $[a,\Pi_{|i}]\ne 0$, because the inclusion of reorientations allows the resulting set of operators to close as an algebra.} This algebra has a rather simple description at the perspective-neutral level. Consider the full algebra $\mathcal{A}_{\mathcal{U}C_i}=\mathcal{A}_{\mathcal{U}} \otimes \mathcal{B}(\mathcal{H}_{C_i})$ of kinematical operators acting on the fields in $\mathcal{U}$, and on frame $C_i$. It then turns out \cite{DVEHK} that the algebra generated by dressed observables and reorientations is just the physical representation $r(\mathcal{A}_{\mathcal{U}C_i}^H)$ of the gauge-invariant subalgebra
\begin{equation}
    \mathcal{A}_{\mathcal{U}C_i}^H = \Big\langle a \in \mathcal{A}_{\mathcal{U}C_i}\mid [a,H_S+H_i]=0\Big\rangle \subseteq\mathcal{A}_{\mathcal{U}C_i}.
\end{equation}
{While $\mathcal{A}_{\mathcal{U}C_i}^H$ always is a von Neumann algebra, we will see in Sec.~\ref{Subsection: N+0} that for $r(\mathcal{A}_{\mathcal{U}C_i}^H)$ it depends on whether at least one other clock exists.} This physical representation can be mapped to an algebra in the perspective of $C_i$ via the reduction map \cite{Hoehn:2019fsy,Hoehn:2020epv,delaHamette:2021oex,DVEHK}:
\begin{equation}
    \mathcal{A}^{\text{phys}}_{\mathcal{U}|i} := \mathcal{R}_i(\tau)r\qty\big(\mathcal{A}_{\mathcal{U}C_i}^H) \mathcal{R}_i(\tau)^\dagger = \Pi_{|i}\mathcal{A}_{\mathcal{U}|i}\Pi_{|i},
    \label{Equation: reduced algebra}
\end{equation}
where $\mathcal{A}_{\mathcal{U}|i}$ is generated by $a\in\mathcal{A}_{\mathcal{U}}$ and (bounded functions of) $H_S+\sum_{j \ne i}H_j$. \

Let us also note that the same physical algebra may be described instead in the perspective of a different clock $C_j\ne C_i$. The result is the original gauge-invariant algebra $\mathcal{A}_{\mathcal{U}C_i}^H$ (an identity factor on $\mathcal{H}_j$ is implicitly dropped) projected onto $\mathcal{H}_{|j}$ with $\Pi_{|j}$:
\begin{equation}
      \mathcal{R}_j(\tau)r\qty\big(\mathcal{A}_{\mathcal{U}C_i}^H) \mathcal{R}_j(\tau)^\dagger = \Pi_{|j}\mathcal{A}_{\mathcal{U}C_i}^H.\label{wrongQRF}
\end{equation}
This will be useful in Sec.~\ref{Subsection: N+0}. {Note that $\Pi_{|j}$ and $\mathcal{A}_{\mathcal{U}C_i}^H$ commute.}

\subsection{PW = CLPW}

Let us briefly show that the coinvariant procedure to implement the constraint invoked by CLPW \cite[Sec.~4.2]{Chandrasekaran:2022cip} (and adopted by \cite{Jensen:2023yxy}) is equivalent to a PW reduction and thereby to an internal QRF perspective (see \cite{DVEHK} for further details). Adapted to our conventions and notation, CLPW consider $\mathcal{U}$ to be the de Sitter static patch associated with an observer carrying a clock $C_1$ and $\mathcal{U}'$ the complementary static patch where another observer carries a clock $C_2$. Initially, both clocks are ideal and so the constraint reads $H=H_S+H_1-H_2$ with $\sigma_i=\mathbb{R}$, where the minus sign highlights that the boost time in $\mathcal{U}'$ runs backward, cf.~Fig.~\ref{Figure: observers in subregion}. This constraint is imposed via the map~\cite[Eq.~(4.10)]{Chandrasekaran:2022cip}
\begin{equation}
    T\ket{\phi}=\int_{\mathbb{R}}\dd{t} e^{-i(H_S+H_1)t}\bra{t}_2\ket{\phi}.
\end{equation}
Comparing with Eq.~\eqref{PWred} and recalling the covariance~\eqref{cstatecov}, it is clear that 
\begin{equation}
    T\ket{\phi}=\mathcal{R}_2(0)\Ket{\phi}=\ket*{\phi_{|2}(0)}\in\mathcal{H}_{|2}.
\end{equation}
In words, CLPW's description of physical states is tantamount to ``jumping into the perspective of the clock QRF $C_2$ when it reads $\tau_2=0$'' via a Page-Wootters reduction. 
Similarly, their description of the physical representation of the static patch algebras $r(\mathcal{A}_{\mathcal{U}C_1}^H)$ and $r(\mathcal{A}_{\mathcal{U}'C_2}^H)$ in \cite[Sec.~4.2]{Chandrasekaran:2022cip} coincides with their PW reduction via Eqs.~\eqref{Equation: reduced algebra} and~\eqref{wrongQRF} into $C_2$ perspective.\footnote{In fact, in $C_2$ perspective, CLPW perform another  unitary to put the crossed product $r(\mathcal{A}_{\mathcal{U}C_1}^H)$ into a standard form.}
While initially, the projectors $\Pi_1,\Pi_2$ are trivial because the clocks are ideal, CLPW later impose lower bounds on their energies and this is equivalent to starting like us with non-ideal clocks and non-trivial projectors from the outset.\footnote{However, this difference does have non-trivial implications for the density operator \cite{DVEHK}.}

Thus, CLPW's observers are nothing but QRFs and their description of states and observables is equivalent to the perspective-neutral QRF formalism. Invoking the latter permits us to generalise their construction (and its generalisation to general subregions $\mathcal{U}$ \cite{Jensen:2023yxy}) to an arbitrary number of observers and to explore the observer dependence of gravitational entropy. This will extend the recent observation that in gauge systems entropies of subsystems are QRF-dependent \cite{Hoehn:2023ehz}. 

\section{The entropy of the subsystem}
\label{Section: entropy}

Let us address now the main topic of the paper: computing the entropy of a subsystem associated with the spacetime subregion $\mathcal{U}$. A subsystem can be described in terms of its algebras of observables, and in Table~\ref{Table: different algebras}, we have summarised the various algebras of physical observables one can use to characterise the degrees of freedom in $\mathcal{U}$. We have given the form of each algebra at both the perspective-neutral level (i.e.\ as algebras acting on $\mathcal{H}_{\text{phys}}$) and in the perspective of $C_i$ (i.e.\ as algebras acting on $\mathcal{H}_{|i}$). As stated earlier, the algebras on any one row of the table are isomorphic via the reduction map.

\begin{table}
    \centering
    \begin{tabular}{lcc}\toprule
        & Perspective-neutral & $C_i$'s perspective \\ \midrule
        Pure QFT & $\mathbb{C}\mathds{1}$& $\mathbb{C}\mathds{1}$\\\addlinespace
        Dressed to $C_i$ & $\mathcal{O}_{\mathcal{U}|i}$ & $\Pi_{|i}\mathcal{A}^{\Pi_{|i}}_{\mathcal{U}}$ \\\addlinespace
        Dressed observables and reorientations & $r(\mathcal{A}_{\mathcal{U}C_i}^H)$ & $\Pi_{|i}\mathcal{A}_{\mathcal{U}|i}\Pi_{|i}$\\ \bottomrule
    \end{tabular}
    \caption{The various observable algebras one can use to characterise the spacetime region $\mathcal{U}$. {When $C_i$ is ideal, and there is at least one other ideal clock in the system, the last row is a crossed product algebra (otherwise a subalgebra of it)}.}
    \label{Table: different algebras}
\end{table}

The algebras in the first row are completely trivial when the modular flow is ergodic, which is often the case as described above. The algebras on the second row are typically Type III {and thus admit neither traces nor density operators}.\footnote{This is easiest to see when $C_i$ is ideal, in which case the algebra in the perspective of $C_i$ is just $\mathcal{A}_{\mathcal{U}}$, which, being the algebra of QFT degrees of freedom in a subregion, is Type $\rm{III}_1$~\cite{10.1143/PTP.32.956,WittenRevModPhys.90.045003,Yngvason_2005}. The non-ideal case will be addressed in~\cite{DVEHK}.} For this reason, we cannot use the algebras in the first two rows to compute a meaningful entanglement entropy. This bears repeating: in the gravitational system we are considering, one cannot compute a well-defined entropy using purely QFT degrees of freedom in $\mathcal{U}$, and it \emph{also} does not suffice to also just allow QFT observables dressed to a clock.

As we will discuss, by going to the third row, i.e.\ by including also clock reorientations, one can make progress. The reason this works was the essential underlying mathematical point of~\cite{Chandrasekaran:2022cip} and the follow-up works {\cite{Jensen:2023yxy,Kudler-Flam:2023hkl,Kudler-Flam:2023qfl,Witten:2023xze}}: the algebra of dressed observables and reorientations serves as a representation of the `crossed algebra' $\mathcal{A}_{\mathcal{U}}\rtimes_\alpha \RR$, where $\alpha:\RR\to \operatorname{Aut}(\mathcal{A}_{\mathcal{U}})$, is the one-parameter family of automorphisms $\alpha_t(\cdot) = e^{-iH_St}(\cdot)e^{iH_St}$. Recalling that $H_S$ generates the modular flow of a state $\ket{\psi_S}$ with respect to the Type III$_1$ algebra $\mathcal{A}_{\mathcal{U}}$, it is a standard mathematical result that the crossed algebra is {a factor} of Type $\mathrm{II}_\infty$~\cite{Takesaki1979}. Thus, one can use it to define traces and entanglement entropies, and this structure descends to its representation $r(\mathcal{A}_{\mathcal{U}C_i}^H)$. We will see in \cite{DVEHK} that the representation $r$ is always faithful, provided there exists at least one other clock. For the special case that there exists another ideal clock $C_j\neq C_i$ this can be easily seen from the right hand side of Eq.~\eqref{wrongQRF}, which implies $r(\mathcal{A}_{\mathcal{U}C_i}^H)\simeq\mathcal{A}^H_{\mathcal{U}C_i}$ since $\Pi_{|j}=\mathds1$, but this holds more generally. If $C_i$ is an ideal clock, we have $\mathcal{A}_{\mathcal{U}C_i}^H\simeq\mathcal{A}_\mathcal{U}\rtimes_\alpha\RR$, and then $r(\mathcal{A}_{\mathcal{U}C_i}^H)$ {also} remains of Type $\mathrm{II}_\infty$. Moreover, if $C_i$ is a non-ideal clock with a lower bound on its energy, $\mathcal{A}_{\mathcal{U}C_i}^H$ is of Type $\rm{II}_1$ and this is also true for the represented algebra $r(\mathcal{A}_{\mathcal{U}C_i}^H)$ as long as another clock exists. Physically speaking, this is due to the existence of a maximally mixed state on this algebra.

A subtlety which is worth noting is as follows. In the above, and elsewhere in the literature, the algebras in question are often assumed to be von Neumann algebras. However, this is not always the case. Von Neumann algebras are required to be equal to their own bicommutant, but the bicommutant of an algebra is determined by the Hilbert space on which it acts. In this way, it may be that an algebra $\mathcal{M}\subset \mathcal{B}(\mathcal{H}_{\text{kin}})$ of gauge-invariant kinematical operators is equal to its bicommutant when acting on the \emph{kinematical} Hilbert space, while its physical representation $r(\mathcal{M})$ is not equal to its own bicommutant when acting on the \emph{physical} Hilbert space. We will discuss this subtlety later --- the punchline is that these problems go away for the case at hand, if there is at least one other clock than $C_i$ in the system, i.e.\ if $n>1$. Then $r(\mathcal{A}_{\mathcal{U}C_i}^H)$ is a von Neumann algebra on the physical Hilbert space. However, if $n=1$, then it is not a von Neumann algebra, and we should replace it by its bicommutant $r(\mathcal{A}_{\mathcal{U}C_i}^H)''$ in order to classify its type, and to compute an entropy.\footnote{{It turns out that the trace on $\mathcal{A}_{\mathcal{U}C_i}^H$ descends to one on $r(\mathcal{A}_{\mathcal{U}C_i}^H)$ even when the latter is only a $C^*$-algebra but not a von Neumann algebra. In this manner, one can in principle still compute density operators and entropies. However, their interpretation will be less transparent as Haag duality \cite{haag2012local,WittenRevModPhys.90.045003} will not hold for this physical algebra. Associating its entropy with the causal diamond $\mathcal{U}$ thus becomes less immediate.}} We will assume for now that $n>1$, postponing further discussion of the $n=1$ case until Sec.~\ref{Subsection: N+0}.

\subsection{The density operator}

Let us now summarise how one can compute the density operator of an arbitrary given physical state $\Ket{\phi}$ with respect to the algebra $r(\mathcal{A}_{\mathcal{U}C_i}^H)$ of dressed observables and reorientations. {(The generalisation to arbitrarily many clocks is saved for later.)} We proceed in several steps.

First, we write down the trace for the gauge-invariant kinematical algebra $\mathcal{A}_{\mathcal{U}C_i}^H$ (in all cases each of the algebras we are dealing with is a factor, and so have a unique trace up to multiplication by a state-independent constant \cite{Sorce:2023fdx}):
\begin{equation}
    \Tr(\cdot) = e^{S_{0,i}}\bra{\psi_S}\bra{0}_i (\cdot) e^{-\beta H_i} \ket{\psi_S}\ket{0}_i.
    \label{Equation: trace}
\end{equation}
Here, {$\ket{0}_i$ is the $t=0$ clock state \eqref{clockstate} and} $e^{S_{0,i}}$ is some arbitrarily chosen normalisation constant, with $S_{0,i}\in\RR$ (we will give a proper interpretation of this constant later below). This functional satisfies all of the required properties of a trace: faithfulness, normality, semifiniteness and the cyclic property (for more on these, see the companion paper~\cite{DVEHK}, or~\cite{Takesaki1979,Witten:2021unn,Chandrasekaran:2022cip,Jensen:2023yxy,Sorce:2023fdx} among many others). It may be thought of as a state on the algebra in which the clock $C_i$ is in thermal equilibrium with the fields in the KMS state $\ket{\psi_S}$.

With this trace one can define the density operator for a physical state on $\mathcal{A}_{\mathcal{U}C_i}^H$. That is, given a $\Ket{\phi}\in\mathcal{H}_{\text{phys}}$, one can define an operator $\rho_\phi\in\mathcal{A}_{\mathcal{U}C_i}^H$ such that\footnote{Note that, thanks to the faithfulness of the trace, $\rho_\phi$ is uniquely defined regardless of whether $r$ is faithful.}
\begin{equation}
    \Tr(\rho_\phi a) = \Mel{\phi}{r(a)}{\phi} \quad \text{for all }a\in\mathcal{A}_{\mathcal{U}C_i}^H.
\end{equation}
The following convenient formula\footnote{We provide a full derivation of this formula in the companion paper~\cite{DVEHK}.} for $\rho_\phi$, which holds for any $\tau$, may be confirmed by direct substitution into the above\footnote{Showing that this is indeed an operator in $\mathcal{A}^H_{\mathcal{U}C_i}$ requires the following fact. Suppose $\mathcal{M}$ is a von Neumann algebra acting on a Hilbert space containing states $\ket{\alpha},\ket{\beta},\ket{\gamma}$, with $\ket{\alpha}$ cyclic and separating. Then $\Delta^{-\frac12}_{\alpha}S_{\beta|\alpha}^\dagger S_{\gamma|\alpha}\Delta^{-\frac12}_\alpha$ is an element of $\mathcal{M}$, where $\Delta_\alpha$ is the modular operator of $\alpha$, and $S_{\beta|\alpha},S_{\gamma|\alpha}$ are the relative Tomita operators from $\ket{\alpha}$ to $\ket{\beta},\ket{\gamma}$ respectively (with respect to $\mathcal{M}$). This may be shown using, for example, the formulas in~\cite[App. C]{Jensen:2023yxy}, and it is also explicitly demonstrated in~\cite{DVEHK}.}:
\begin{equation}
    \rho_\phi = e^{\beta (H_S+H_i)-S_{0,i}} \int_{\mathbb{R}} \dd{t} e^{-iH_it} O^\tau_{C_i}(S^\dagger_{\phi_{|i}(\tau+t)|\psi_S}S_{\phi_{|i}(\tau)|\psi_S}),
\end{equation}
where $S_{\phi_{|i}(t)|\psi_S}$ is the relative Tomita operator between $\ket{\psi_S}$ and the state $\ket*{\phi_{|i}(t)}$ of the system in the perspective of $C_i$,\footnote{These are a part of Tomita-Takesaki theory, which is the mathematical formalism most commonly used for understanding the entanglement properties of Type III algebras like $\mathcal{A}_{\mathcal{U}}$~\cite{WittenRevModPhys.90.045003}.} for the algebra $\mathcal{A}_{\mathcal{U}}$, defined by
\begin{equation}
    S_{\phi_{|i}(t)|\psi_S}a\ket{\psi_S} = a^\dagger \ket*{\phi_{|i}(t)} = a^\dagger \mathcal{R}_i(t)\Ket{\phi}.
\end{equation}
In this way, our formula links the density matrix of the physical state $\Ket{\phi}$ with the modular properties of the QFT in the perspective of $C_i$.

Next, we find the density operator of the state $\Ket{\phi}$ in the \emph{physical} algebra $r(\mathcal{A}_{\mathcal{U}C_i}^H)$. To do so, we show in the companion paper~\cite{DVEHK} that the trace defined on $\mathcal{A}_{\mathcal{U}C_i}^H$ leads to a trace on its physical representation $r(\mathcal{A}_{\mathcal{U}C_i}^H)$, and furthermore that the corresponding physical density operator is none other than the physical representation of $\rho_\phi$: 
\begin{equation}
    \rho_\phi^{\text{phys}} := r(\rho_\phi) = r\qty(e^{\beta H_S}) e^{-S_{0,i}} \int_{\mathbb{R}}\dd{t} V_{C_i}(t+i\beta) \mathcal{O}^\tau_{C_i}(S^\dagger_{\phi_{|i}(\tau+t)|\psi_S}S_{\phi_{|i}(\tau)|\psi_S}).
\end{equation}
We can map this into an operator in $\mathcal{A}_{\mathcal{U}|i}^{\text{phys}}$ (i.e.\ the corresponding algebra in the perspective of $C_i$, defined in~\eqref{Equation: reduced algebra}) using the isomorphism between the perspective-neutral and reduced Hilbert spaces:
\begin{align}
    \rho_{\phi|i}(\tau) &= \mathcal{R}_i(\tau)\rho_\phi^{\text{phys}}\mathcal{R}_i(\tau)^\dagger\\
    &= \Pi_{|i}e^{-\beta \sum_{j\ne i} H_j-S_{0,i}} \int_{\mathbb{R}}\dd{t} e^{i(H_S+\sum_{j\ne i}H_j)t} S^\dagger_{\phi_{|i}(\tau+t)|\psi_S}S_{\phi_{|i}(\tau)|\psi_S}.
\end{align}
{It turns out that the first term $\Pi_{|i}$ commutes with the rest of this expression.} We should point out at this stage the conceptual usefulness of this formula: everything now is expressed in the perspective of $C_i$. There is therefore a self-contained story which can be told entirely from the perspective of the clock. It should also be noted that the reduced perspective density operator obeys the relational Liouville equation
\begin{equation}
    \rho_{\phi|i}(\tau') = e^{-i(H_S+\sum_{j\ne i} H_j)(\tau'-\tau)} \rho_{\phi|i}(\tau)e^{i(H_S+\sum_{j\ne i} H_j)(\tau'-\tau)}.
\end{equation}
Thus, this story is consistent with the relational quantum dynamics picture alluded to earlier {and in particular the PW formalism.}

As we will show in the companion paper~\cite{DVEHK}, our formula for the density operator recovers the more specialised cases investigated in~\cite{Chandrasekaran:2022cip,Jensen:2023yxy}  under the assumptions made there.  In fact, it is a generalisation of the construction in \cite{Chandrasekaran:2022cip,Jensen:2023yxy} which was applied to product states (and projections thereof with $\Pi_{|i}$), whereas our formula holds for arbitrary global physical states.

Armed with the density operator of the state $\Ket{\phi}$, we can now in principle find its entanglement entropy with
\begin{equation}
    S[\phi] = -\Mel{\phi}{\log\rho_\phi^{\text{phys}}}{\phi} = -\mel*{\phi_{|i}(\tau)}{\log \rho_{\phi|i}(\tau)}{\phi_{|i}(\tau)}.
\end{equation}
In practice, however, the logarithm of the density operator is not simple to compute. Later, we will describe a semiclassical regime in which the calculation simplifies, as well as a few examples that fall outside of this regime.

\subsection{Normalisation and the type of the algebra}
\label{Subsection: normalisation and type}

Let us comment on the meaning of the constant $S_{0,i}\in\RR$ that went into the definition of the trace in Eq.~\eqref{Equation: trace}. The density operator of a state, and its entropy, depend on this constant. Indeed, if we were to change $S_{0,i}\to S_{0,i}'$, then the density operators given above would also change by a factor $\exp(S_{0,i}-S_{0,i}')$, and the entropy of any state $\Ket{\phi}$ would change via
\begin{equation}
    S[\phi] \to S[\phi] + S_{0,i}'-S_{0,i}.
\end{equation}
Thus, the freedom to choose $S_{0,i}$ reflects an overall (state-independent) ambiguity in the definition of the entropy. Note that $S[\phi_1]-S[\phi_2]$ is invariant under $S_{0,i}\to S_{0,i}'$, for any two states $\Ket{\phi_1},\Ket{\phi_2}$, so the difference in entropies of two states is completely unambiguous.

There are various ways to fix the constant $S_{0,i}$, and so remove the ambiguity entirely. One way is to pick a particular reference state, and choose $S_{0,i}$ such that the entropy of the reference state vanishes. For example, if the trace defined above is \emph{finite}, meaning $\Tr(\mathds{1})$ converges, then the ambiguity in $S_{0,i}$ can be fixed by setting
\begin{equation}\label{normal1}
    \Tr(\mathds{1}) = e^{S_{0,i}} \bra{0}_ie^{-\beta H_i}\ket{0}_i = 1,
\end{equation}
i.e.\ 
\begin{equation}\label{normal2}
    S_{0,i} = -\log \frac{Z_i}{2\pi}, \qq{where} Z_i := 2\pi\bra{0}_ie^{-\beta H_i}\ket{0}_i = \int_{\sigma_i}e^{-\beta \varepsilon}\dd{\varepsilon}.
\end{equation}
Note that the $\varepsilon$-integral in $Z_i$ needs to converge, which means there must be a lower bound on the energy spectrum $\sigma_i$ of the clock. With this normalisation, the maximally mixed state (i.e.\ the state with maximal entropy) on the gauge-invariant algebra $\mathcal{A}_{\mathcal{U}C_i}^H$ has vanishing entropy (indeed, the density operator of this state is just the identity $\mathds{1}$, and $-\Tr(\mathds{1}\log\mathds{1})=0$). This is what was done in~\cite{Chandrasekaran:2022cip}.

A point of subtlety here is that the maximally mixed state on the gauge-invariant \emph{kinematical} algebra $\mathcal{A}_{\mathcal{U}C_i}^H$ may not in fact correspond to a physical state of the system. Indeed, given a state on the \emph{physical} algebra $r(\mathcal{A}_{\mathcal{U}C_i}^H)$, which is in general a functional $\Psi:r(\mathcal{A}_{\mathcal{U}C_i}^H)\to \CC$, the corresponding state on the gauge-invariant algebra is given by $\Psi_r: a \mapsto \Psi(r(a))$. But it may be the case that the maximally mixed state on $\mathcal{A}_{\mathcal{U}C_i}^H$ cannot be written as $\Psi_r$ for any choice of $\Psi$. The maximally mixed state on the physical algebra will typically have a smaller entropy than the maximally mixed state on the gauge-invariant algebra, because there are fewer degrees of freedom at play after imposing the constraint $H=0$. 

{More generally, since the trace defined on the gauge-invariant kinematical algebra $\mathcal{A}_{\mathcal{U}C_i}^H$ descends to one on its physical representation $r(\mathcal{A}_{\mathcal{U}C_i}^H)$ \cite{DVEHK}, this means that fixing the normalisation for the kinematical trace via Eqs.~\eqref{normal1} and~\eqref{normal2} also fixes the normalisation of the \emph{physical} trace for $r(\mathcal{A}_{\mathcal{U}C_i}^H)$. Since traces in von Neumann algebras are unique up to scaling (e.g., see \cite{Sorce:2023fdx}), this fixes the entropy of the maximal entropy state when the algebra is of Type $\rm{II}_1$, and the previous paragraph indicates that this may not be zero if normalised kinematically. Alternatively, one can directly normalise the trace at the physical level to remedy this. Further discussion of the subtleties and ambiguities involved in normalising the traces of the various algebras may be found in~\cite{DVEHK}.}

In the case of an ideal clock $C_i$, the dressed operators at $\tau=0$ may be written \cite{Hoehn:2019fsy}
\begin{equation}
    O_{C_i}^0(a) =e^{-iH_S\hat{t}_i}a e^{iH_S\hat{t}_i},\qq{where}
    \hat{t}_i := \int_{\mathbb{R}}\dd{t} t\dyad{t}_i.
\end{equation}
As explained in~\cite{Chandrasekaran:2022cip}, the algebra generated by these operators and $H_i$, which we have been calling $\mathcal{A}_{\mathcal{U}C_i}^H$, is the crossed algebra $\mathcal{A}_{\mathcal{U}}\rtimes_\alpha\RR$, where $\alpha$ is the modular flow of $\ket{\psi_S}$. It is a mathematical result that this algebra is Type $\mathrm{II}_\infty$ \cite{Takesaki1979}. The algebra {$\mathcal{A}_{\mathcal{U}C_i}^H$} in the case of a non-ideal $C_i$ can be understood as a subalgebra of this crossed algebra, in which we project onto states with the energy of the clock falling in the spectrum $\sigma_i$; this subalgebra is also a Type II factor (it is sometimes called a `corner' or `compression' of the crossed product). The physical algebra $r(\mathcal{A}_{\mathcal{U}C_i}^H)$ is thus overall a representation of a Type II subalgebra of the crossed algebra $\mathcal{A}_{\mathcal{U}}\rtimes_\alpha \RR$. Any non-trivial representation of a Type II algebra that preserves the von Neumann property (i.e.\ a normal representation) is also Type II,\footnote{This can be seen by noting that a Type II algebra is an algebra with a trace but no non-trivial irreducible representations.} so $r(\mathcal{A}_{\mathcal{U}C_i}^H)$ is Type II (assuming again that there are $n>1$ clocks); moreover, since $r$ is faithful (see \cite{DVEHK}), $r(\mathcal{A}_{\mathcal{U}C_i}^H)$ is a factor. When the trace is finite, which by Eq.~\eqref{normal2} occurs when $\sigma_i$ is bounded below, it is Type $\mathrm{II}_1$; otherwise it is Type $\mathrm{II}_\infty$.

\subsection{Using multiple frames to observe the QFT}

So far, we have been considering the use of a single clock $C_i$ in the algebra of observables for the subregion $\mathcal{U}$. However, there are many clocks available in the system we are considering, and it is reasonable to ask what can be gained by using several of them at the same time. There is a natural generalisation of the previous structure to this case.

Indeed, suppose $R\subseteq \{C_1,\dots,C_n\}$ is some subset of the QRFs, and let $R^c=\{C_1,\dots,C_n\}\setminus R$ be its complement.\footnote{Here, by complement of $R$, we are not referring in any way to the causal structure of spacetime -- we just mean the set of clocks that are not in $R$. {(It thus refers to the kinematical subsystem complement of $R$ on $\mathcal{H}_{\rm kin}$ in Eq.~\eqref{Hkin}.)} Indeed, the clocks in $R$ can be located in either $\mathcal{U}$ or its causal complement $\mathcal{U}'$, and the same is true of $R^c$; it makes no difference for the arguments we will make.} Suppose we pick a particular clock $C_i\in R$. Then we may view the fields in $\mathcal{U}$, and the clocks in $R_{\bar{i}}=R\setminus \{C_i\}$, as the collection of degrees of freedom that the clock $C_i$ is being used to observe. Thus, we can dress observables 
\begin{equation}
    a\in \mathcal{A}_{\mathcal{U}}\otimes\bigotimes_{C_j\in R_{\bar{i}}} \mathcal{B}(\mathcal{H}_j)
\end{equation}
to the clock $C_i$ to obtain 
\begin{equation}
    O^{\tau}_{C_i}(a) =  \int_{\mathbb{R}}\dd{t}e^{-i\qty(H_S + \sum_{C_j\in R} H_j)t}\qty(a \otimes \dyad{\tau}_i) e^{i\qty(H_S + \sum_{C_j\in R} H_j)t}.
\end{equation}
The physical representations of these dressed observables, along with the reorientations of $C_i$, generate the physical representation $r(\mathcal{A}_{\mathcal{U}R}^H)$ of the {von Neumann} algebra 
\begin{equation}
    \mathcal{A}_{\mathcal{U}R}^H = \Big\langle a\in \mathcal{A}_{\mathcal{U}}\otimes \bigotimes_{C_i\in R}\mathcal{B}(\mathcal{H}_{i})\mid [a,H_S + \sum_{C_i\in R} H_i] = 0\Big\rangle
\end{equation}
of gauge-invariant observables of the fields in $\mathcal{U}$ and the clocks in $R$.

We presented this algebra in terms of a particular clock $C_i$, but we clearly get the same algebra for each possible choice of $C_i\in R$. An alternative construction involves considering observables dressed to all of the clocks in $R$ simultaneously; along with the reorientations of each clock in $R$, these generate the same algebra $r(\mathcal{A}_{\mathcal{U}R}^H)$. The algebra thus describes the set of degrees of freedom in $\mathcal{U}$ accessible using all of the clocks in $R$.

When $R^c$ is non-empty, $r(\mathcal{A}_{\mathcal{U}R}^H)$ is {also} a von Neumann algebra~\cite{DVEHK}. We focus here on this case, reserving the case in which $R^c$ is empty for Sec.~\ref{Subsection: N+0}. 

The gauge-invariant kinematical algebra $\mathcal{A}_{\mathcal{U}R}^H$ admits a trace:
\begin{equation}
    \Tr(\cdot) = e^{S_{0,R}}\bra*{\psi_{SR_{\bar{i}}}}\bra{0}_i (\cdot) e^{-\beta H_i} \ket*{\psi_{SR_{\bar{i}}}}\ket{0}_i,
\end{equation}
where $\ket*{\psi_{SR_{\bar{i}}}} = \ket*{\psi_S}\otimes\ket*{\psi_{R_{\bar{i}}}}$, and $\ket*{\psi_{R_{\bar{i}}}}$ is an (unnormalised)  thermofield double state at inverse temperature $\beta$ for the frames in $R_{\bar{i}}$:
\begin{equation}
    \ket*{\psi_{R_{\bar{i}}}} = \bigotimes_{j\in R_{\bar{i}}} \int_{\sigma_j} e^{-\beta \varepsilon/2} \ket{\varepsilon}_j\ket{\varepsilon}_j\dd{\varepsilon},
    \label{Equation: thermofield double}
\end{equation}
where $\ket{\varepsilon}_j$ are eigenstates of $H_j$.\footnote{This state is used for the following reasons. $\ket*{\psi_{R_{\bar{i}}}}$ is cyclic and separating for $\bigotimes_{C_j\in R_{\bar i}}\mathcal{B}(\mathcal{H}_j)$, and the modular flow of $\ket*{\psi_{SR_{\bar{i}}}}$ {on $\mathcal{A}_{\mathcal{U}}\otimes\bigotimes_{C_j\in R_{\bar{i}}}\mathcal{B}(\mathcal{H}_{j})$ agrees with that of} gauge transformations. These properties make it simple to generalise the single-clock approach.} The meaning of the constant $S_{0,R}\in\RR$ will be explained shortly.  

Using this trace we can define density operators and entanglement entropies. In particular, the physical density operator corresponding to an arbitrary state $\Ket{\phi}$ is given at the perspective-neutral level by
\begin{equation}
    \rho_\phi^{\text{phys}} = e^{-S_{0,R}}V_i(i\beta/2) \int_{\mathbb{R}} \dd{t} V_i(t) \mathcal{O}^\tau_{C_i}\qty\Big(\Delta_{\psi_{SR_{\bar{i}}}}^{-1/2}S^\dagger_{\phi_{|i}(\tau+t)|\psi_{SR_{\bar{i}}}}S_{\phi_{|i}(\tau)|\psi_{SR_{\bar{i}}}}\Delta_{\psi_{SR_{\bar{i}}}}^{-1/2})V_i(i\beta/2) .\label{rhophys1}
\end{equation}
and, in the perspective of any $C_i\in R$, it is given by
\begin{equation}
    \rho_{\phi|i}(\tau) = \Pi_{|i}e^{-\beta \sum_{C_j\in R^c} H_j-S_{0,R}}e^{\beta \sum_{C_j\in R_{\bar{i}}} \tilde H_j}\int_{-\infty}^{\infty} \dd{t} e^{i(H_S+\sum_{j\ne i}H_j)t} S^\dagger_{\phi_{|i}(\tau+t)|\psi_{SR_{\bar{i}}}}S_{\phi_{|i}(\tau)|\psi_{SR_{\bar{i}}}}.\label{rhored1}
\end{equation}
Here, the relative Tomita operators are for the algebra $\mathcal{A}_{\mathcal{U}}\otimes\bigotimes_{C_j\in R_{\bar{i}}}\mathcal{B}(\mathcal{H}_{j})$ (which we take to act only on one of the copies of the frames in the thermofield double), and $\Delta_{\psi_{SR_{\bar{i}}}}$ is the modular operator of $\ket*{\psi_{SR_{\bar{i}}}}$ with respect to this same algebra. Also $\tilde{H}_j$ is the Hamiltonian of the copy of $C_j$ in the thermofield double (N.B.\ there is no overall action of $\rho_{\phi|i}(\tau)$ on the copied frames -- this is confirmed by Eq.~\eqref{Equation: intro split up density operator}). These expressions work for any choice of $C_i\in R$, and simplify to the previous ones when $R$ contains only a single frame. 

The gauge-invariant algebra $\mathcal{A}^H_{\mathcal{U}R}$ is {a} Type II factor \cite{DVEHK}. Moreover, if $R$ contains more than one clock, then $\mathcal{A}^H_{\mathcal{U}R}$ is Type $\mathrm{II}_\infty$. This is because $\Tr(\mathds{1})$ is proportional to $\braket*{\psi_{R_{\bar{i}}}}{\psi_{R_{\bar{i}}}}$, where $\ket*{\psi_{R_{\bar{i}}}}$ is the thermofield double state defined in Eq.~\eqref{Equation: thermofield double} -- but this state has infinite norm.\footnote{This has happened because the clocks we are using have infinite-dimensional Hilbert spaces, and $\mathcal{A}_{\mathcal{U}R}^H$ contains subalgebras such as $\left(\mathcal{B}(\mathcal{H}_1)\otimes\mathcal{B}(\mathcal{H}_2)\right)^{H_1+H_2}$, which are direct sums of type $\rm{I}_\infty$ factors (with the centre generated by $H_1+H_2$). If one were to use clocks with finite-dimensional Hilbert spaces such as in~\cite{periodic}, one could have a maximally mixed state.} The constant $S_{0,R}$ represents an overall state-independent additive ambiguity in the entropy, which for this reason cannot be fixed by requiring the maximally mixed state on the gauge-invariant algebra to have zero entropy (since such a maximally mixed state does not exist). One must fix it by other means, as discussed in~\cite{DVEHK}. For the purposes of this paper we will leave it undetermined.

At this point, the entropies corresponding to these density operators may be computed, but before moving on to actual entropy calculations, it is worth refining the above slightly in order to distinguish the contributions of the QFT and the frames. To that end, suppose the state in the perspective of $C_i\in R$ has the bipartite decomposition
\begin{equation}\label{Equation: Schmidt Decomposition}
    \ket*{\phi_{|i}(t)} = \mathcal{R}_i(t)\Ket{\phi} = \sum_I \ket*{\phi_S^I(t)} \otimes \ket*{\tilde\phi^I(t)},
\end{equation}
where $\ket*{\phi_S^I}\in\mathcal{H}_S$ are a set of not necessarily orthonormal states for the QFT, $\ket*{\tilde\phi^I}\in\bigotimes_{j\ne i}\mathcal{H}_{j}$ are similarly a set of not necessarily orthonormal states for all the frames excluding $C_i$, and $I$ could be either a discrete or a continuous index. One then may write the density operators given above in terms of this decomposition:
\begin{multline}
    \rho_{\phi|i}(\tau) = \Pi_{|i}e^{-\beta \sum_{C_j\in R^c} H_j-S_{0,R}}\\\sum_{I,J}\int_{-\infty}^{\infty} \dd{t} e^{i(H_S+\sum_{j\ne i}H_j)t} 
    S^\dagger_{\phi^I_S(\tau+t)|\psi_S}S_{\phi^J_S(\tau)|\psi_S} \otimes \tr_{R^c}\qty\big(\ket*{\tilde\phi^I(\tau+t)}\!\!\bra*{\tilde\phi^J(\tau)}).
    \label{Equation: intro split up density operator}
\end{multline}
Here, $\tr_{R^c}$ is the standard matrix trace on $\bigotimes_{j\in R^c}\mathcal{B}(\mathcal{H}_{j})$ (used as a partial trace), and the relative Tomita operators are for the algebra $\mathcal{A}_{\mathcal{U}}$.

\subsection{Entropy in a semiclassical regime} \label{sect:semiclassical}

Consider the general case in which one can use multiple clocks, which is captured by the algebra $r(\mathcal{A}_{\mathcal{U}R}^H)$ (if we want to just use a single clock $C_i$, we can set $R=\{C_i\}$). As we have stated above, the entanglement entropy with respect to this algebra of a general state $\Ket{\phi}$ is difficult to compute in practice, because the logarithm of the density operators found above will have a very complicated form containing non-commuting operators. To make progress, one needs to consider a restricted class of states. In particular, we will consider a generalisation of the semiclassical regime discussed by~\cite{Chandrasekaran:2022cip,Jensen:2023yxy}, which was based on a condition in which the wavefunction of a clock is slowly varying in an energy eigenbasis. Our description is conversely based on a sharp peaking in time, which{, while encompassing the states of \cite{Chandrasekaran:2022cip,Jensen:2023yxy},} is perhaps more intuitive, and also potentially easier to generalise to the case of QRFs transforming under more complicated groups than just $\RR$. 
 
For the purposes of this overview note, let us consider the case where the reduced state in the perspective of $C_i$ is well approximated by a product state between the QFT and the rest of the frames:\footnote{Note that $\ket*{\phi_{|i}(t)}$ must be in the image of $\Pi_{|i}$, restricting which kinds of product states are allowed here.}
\begin{equation}
    \ket*{\phi_{|i}(t)} \approx \ket*{\phi_S(t)}\otimes\ket*{\tilde\phi(t)}.
    \label{eq:productState}
\end{equation}
Thus, we can remove the indices $I,J$ in Eq.~\eqref{Equation: intro split up density operator}. In the companion paper~\cite{DVEHK}, we consider also states where this is not the case, i.e.\ where there is non-negligible entanglement, which is important because even though the QFT and frames may appear to be in a product state from the perspective of one clock, they will typically be entangled from the perspectives of other clocks. {This is rooted in a general trade-off between superpositions and entanglement under QRF transformations \cite{Hoehn:2019fsy,Hoehn:2020epv,Giacomini:2017zju,Castro-Ruiz:2019nnl,AliAhmad:2021adn,Hoehn:2023ehz,delaHamette:2020dyi}. In fact, in Sec.~\ref{sec_interferometer}, we will illustrate a gravitational interferometer example of this and how this renders the entropy clock-dependent.}

To get to a semiclassical regime, we make two additional assumptions:
\begin{enumerate}
    \item Roughly speaking, the time read by clock $C_i$ is very sharply peaked in a window of $\order{\epsilon}$, for some small $\epsilon$. This can be captured by the following condition on the physical state:
    \begin{equation}
        \abs{\Mel{\phi}{V_i(t)}{\phi}}\ll 1 \qq{if} \abs{t} > \order{\epsilon}.
    \end{equation}
    In words, if we reorient the clock more than $\order{\epsilon}$, we get a state which is almost orthogonal to the one we started with. {In quantum information parlance, this means that we require the state to be highly asymmetric under reorientations, and this necessitates $\Ket{\phi}$ to be highly entangled across the \emph{kinematical} tensor product structure~\eqref{Hkin} \cite{delaHamette:2021piz} (this is thus not a physical entanglement).} We will furthermore assume that this property is not disturbed by acting with operators dressed to the clock:
    \begin{equation}
        \abs{\Mel{\phi}{\mathcal{O}_{C_i}^\tau(a) V_i(t)}{\phi}}\ll \abs{\Mel{\phi}{\mathcal{O}_{C_i}^\tau(a)}{\phi}} \qq{if} \abs{t} > \order{\epsilon},
        \label{Equation: intro semiclassical 1}
    \end{equation}
    for any $a\in\mathcal{A}_{\mathcal{U}}\otimes\bigotimes_{C_j\in R_{\bar{i}}}\mathcal{B}(\mathcal{H}_{C_j})$. This condition can only hold if there is \emph{some} physical degree of freedom (or collection of degrees of freedom) in the state $\Ket{\phi}$, almost entirely uncorrelated from the dressed observables, and keeping track of the time read by $C_i$ with very little uncertainty.
    \item In the perspective of $C_i$, the QFT part of the state is approximately constant over times of $\order{\epsilon}$. More precisely, we assume
    \begin{equation}
        \ket{\phi_S(t')}\approx \ket{\phi_S(t)} \qq{if} \abs{t'-t}<\order{\epsilon}.
        \label{Equation: intro semiclassical 2}
    \end{equation}
    In other words, the fields are in an approximate energy eigenstate. For simplicity here, we are taking the energy eigenvalue to be negligible, but in~\cite{DVEHK}, this assumption is relaxed to allow the fields to have non-negligible energy -- the results of this section are unaltered.
\end{enumerate}
It should be clear the sense in which this regime is \emph{semiclassical}. The first assumption provides us with an approximately \emph{classical} time variable (roughly speaking, the time of clock $C_i$), and the second assumption says that at the time given by that variable, there is a fixed \emph{quantum} state for the fields.

It turns out that the sharp peaking described by Eq.~\eqref{Equation: intro semiclassical 1} translates directly to a sharp peaking in the $t$ integral in our expression for the density operator in Eqs.~\eqref{rhophys1} and~\eqref{rhored1}.\footnote{Indeed, suppose $a,b\in\mathcal{A}_{\mathcal{U}}\otimes\bigotimes_{C_j\in R_{\bar{i}}}\mathcal{B}(\mathcal{H}_{j})$. Then we have
\begin{equation}
    \mel*{\psi_{SR_{\bar{i}}}}{b S_{\phi(t+\tau)|\psi_{SR_{\bar{i}}}}^\dagger S_{\phi(\tau)|\psi_{SR_{\bar{i}}}}a}{\psi_{SR_{\bar{i}}}} = \Mel{\phi}{\mathcal{O}_{C_i}^\tau(ab)V_i(t)^\dagger}{\phi},
\end{equation}
and Eq.~\eqref{Equation: intro semiclassical 1} implies that this is suppressed for $\abs{t}>\order{\epsilon}$. But since $\ket*{\psi_{SR_{\bar{i}}}}$ is cyclic for $\mathcal{A}_{\mathcal{U}}\otimes\bigotimes_{C_j\in R_{\bar{i}}}\mathcal{B}(\mathcal{H}_{j})$, it must be true that the operator $S_{\phi(t+\tau)|\psi_{SR_{\bar{i}}}}^\dagger S_{\phi(\tau)|\psi_{SR_{\bar{i}}}}$ itself is suppressed for $\abs{t}>\order{\epsilon}$.} To be precise, the integral is dominated by contributions from $\abs{t}< \order{\epsilon}$. At these peaks, Eq.~\eqref{Equation: intro semiclassical 2} means that the QFT part of the integrand may be treated as constant in $t$, and factored out of the integral. As a result, the density operator takes the following approximate form (in the perspective of $C_i$):
\begin{equation}
    \rho_{\phi|i}(\tau) \approx e^{-S_{0,R}}\Delta_{\phi_S(\tau)|\psi_S} e^{-\beta\sum_{C_i\in R^c}H_i} \rho_{R|i}(\tau).
\end{equation}
Here, $\Delta_{\phi_S(\tau)|\psi_S}$ is the relative modular operator from the QFT {KMS state} $\ket{\psi_S}$ to the state $\ket{\phi_S(\tau)}$ in the perspective of $C_i$, and
\begin{equation}
    \rho_{R|i}(\tau) = \Pi_{|i}\int_{-\infty}^{\infty} \dd{t} e^{i(H_S+\sum_{j\ne i}H_j)t}
    \braket*{\phi_S(\tau)}{\phi_S(\tau+t)} \tr_{R^c}\qty\big(\ket*{\tilde\phi(\tau+t)}\!\!\bra*{\tilde\phi(\tau)})
\end{equation}
has the interpretation of a density operator for the clocks in $R$, from the perspective of $C_i$.\footnote{More precisely, consider the algebra $\mathcal{A}_{R|i}^{\text{phys}} = \Pi_{|i}\mathcal{A}_{R|i}\Pi_{|i}$, where $\mathcal{A}_{R|i}$ is generated by $a\in \bigotimes_{j\in R_{\bar{i}}}\mathcal{B}(\mathcal{H}_{j})$ and $H_S+\sum_{j\ne i}H_j$. This algebra describes the clocks in $R$ from the perspective of $C_i$ (including reorientations of $C_i$). A trace on the algebra can be defined by
\begin{equation}
    \Tr\qty\Big(ae^{i(H_S+\sum_{C_j\in R^c}H_j)t}) = \tr_{R_{\bar{i}}}(a) \delta(t),
\end{equation}
where $a\in \bigotimes_{j\in R_{\bar{i}}}\mathcal{B}(\mathcal{H}_{j})$. Using this trace, one may show that $\rho_{R|i}(\tau)$ is the density operator of the physical state $\Ket{\phi}$ with respect to the algebra $\mathcal{A}_{R|i}^{\text{phys}} = \Pi_{|i}\mathcal{A}_{R|i}\Pi_{|i}$.}

The various terms in the density operator approximately commute with each other, which makes finding a simplified expression for its logarithm, and hence the entropy, relatively straightforward. One finds
\begin{equation}
    S[\phi] \approx S_{0,R} - S_{\text{rel}}(\phi_S(\tau)||\psi_S) - \beta \sum_{C_i\in R}\expval{H_i}_{\phi} + S_R[\phi],
    \label{Equation: intro semiclassical entropy}
\end{equation}
where\footnote{The second equality is only well-defined when $\phi_S$ is separating and requires Connes' cocycle theorem (e.g., see \cite{Jensen:2023yxy}). When $\phi_S$ is not separating, $S_{\text{rel}}(\phi_S(\tau)||\psi_S)$ is infinite \cite{WittenRevModPhys.90.045003}.} 
\begin{align}
    S_{\text{rel}}(\phi_S(\tau)||\psi_S) &= -\bra{\phi_S(\tau)}\log\Delta_{\psi_S|\phi_S(\tau)}\ket{\phi_S(\tau)} \\
    &= \bra{\phi_S(\tau)}\log \Delta_{\phi_S(\tau)|\psi_S}\ket{\phi_S(\tau)} + \beta \mel{\phi_S(\tau)}{H_S}{\phi_S(\tau)}.
\end{align}
is the relative entropy of $\ket{\phi_S(\tau)}$ compared to the {KMS} state $\ket{\psi_S}$, with respect to the algebra $\mathcal{A}_{\mathcal{U}}$~\cite{Araki:1976zv},\footnote{Despite appearances, this term is $\tau$-independent.} and 
\begin{equation}
    S_R[\phi] = -\bra*{\tilde\phi(\tau)}\log\rho_{R|i}(\tau)\ket*{\tilde\phi(\tau)}
\end{equation}
is the entanglement entropy of the clocks in $R$. This generalises a result of~\cite{Chandrasekaran:2022cip} to the case in which one is allowed to dress observables to arbitrarily many clocks. If desired, one may write this in terms of a generalised gravitational entropy (i.e.\ involving an area term), following the procedure described in~\cite{Jensen:2023yxy,DVEHK}. 

Let us now give a bit more detail on when exactly one can expect the semiclassical approximation to work. The sharp peaking in the $t$ integral in Eq.~\eqref{Equation: intro split up density operator} cannot come from the QFT part of the state, by Eq.~\eqref{Equation: intro semiclassical 2}. The only term which can be responsible for it is the part involving the frames, i.e.\ $\tr_{R^c}(\ket*{\tilde{\phi}(\tau+t)}\!\!\bra*{\tilde{\phi}(\tau)})$. In particular, due to the partial trace $\tr_{R^c}$ it is the frames in the \emph{complement} $R^c$ which are responsible for the semiclassical regime (perhaps slightly unintuitively). {Note that $\tr_{R^c}(\ket*{\tilde{\phi}(\tau+t)}\!\!\bra*{\tilde{\phi}(\tau)})$ contains a transition amplitude for the complementary clocks and this must be sharply peaked around $t=0$.} To be specific, the {complementary clocks} must have support over a wide range of energies (this is a gauge-invariant condition). For example, there could be just one clock in the complement, whose wavefunction in the energy eigenbasis is slowly varying (this is essentially the scenario studied in \cite{Chandrasekaran:2022cip,Jensen:2023yxy}, except that those authors reduce with respect to a clock in $\mathcal{U}'$ and so the `slowly varying' property would refer to a clock in $\mathcal{U}$). 
On the other hand, there could be many clocks in $R^c$, each of which has an arbitrarily narrow range of energies. Collectively, these clocks would have a total energy supported over a broad range, and in a ``thermodynamic limit'' in which the number of clocks in the complement becomes very large, by the law of large numbers the term $\tr_{R^c}(\ket*{\tilde{\phi}(\tau+t)}\!\!\bra*{\tilde{\phi}(\tau)})$ would become sharply peaked in $t$; this would therefore give suitable conditions for a semiclassical approximation. It should be noted that this latter example is the natural case in realistic physical situations -- there are, after all, many clocks in the universe!

\subsection{Beyond the semiclassical regime?}\label{sec_beyond}

We show in the companion paper~\cite{DVEHK} that the entropy formula Eq.~\eqref{Equation: intro semiclassical entropy} holds to linear order in the semiclassical approximation. Finding quadratic and higher order corrections to the formula quickly becomes rather complicated, and we will make no attempt here to do so in the general case.

Instead, {we will study an illustrative example in Sec.~\ref{sec_interferometer}, where the product state assumption in Eq.~\eqref{eq:productState} is violated in one perspective, due to the corresponding clock passing through a gravitational interferometer in another. More general entangled states and corrections to Eq.~\eqref{Equation: intro semiclassical entropy}} are also explored in \cite{DVEHK}. 

Another interesting regime is as follows: consider a state $\Ket{\phi}$ in which the energy of a given clock $C_i$ is supported within a very narrow window. In many ways this is the opposite of the semiclassical regime; for example, it leads to $\abs{\Mel{\phi}{V_i(t)}{\phi}}\approx 1$ for a very large range of times $t$, in contrast to Eq.~\eqref{Equation: intro semiclassical 1}. We call this an `antisemiclassical' regime. The leading order contribution to the entropy of $\Ket{\phi}$ with respect to the algebra $r(\mathcal{A}_{\mathcal{U}C_i}^H)$ is completely independent of the QFT part of $\Ket{\phi}$, so long as the energy window of the clock is sufficiently small.\footnote{One may show this by seeing what happens when one computes the expectation value in $\Ket{\phi}$ of an operator $a\in r(\mathcal{A}_{\mathcal{U}C_i}^H)$. This is equivalent to computing the expectation value of $PaP$, where $P$ is a projection onto the range of energies for the clock $C_i$ in the state $\Ket{\phi}$. Since this range is very small, we have $[PaP,H_i]\approx 0$, and since $PaP$ is gauge-invariant we also have $[PaP,H_S]\approx 0$. This implies we can approximate $PaP$ by an operator in $\mathcal{A}_{\mathcal{U}}^H\otimes \mathcal{B}(\mathcal{H}_i)^H=\mathds{1}_S\otimes\mathcal{B}(\mathcal{H}_i)^H$, and the expectation value of such an operator does not depend at all on the QFT part of the state. (Here we are assuming the QFT modular flow is ergodic, which is commonly the case, as explained elsewhere in the paper.)}  The intuition for this is that the peaking in energy causes the time of the clock to have very large fluctuations, which essentially destroys all hopes of using it to gain any information about the field degrees of freedom. We will come back to this below and in more detail in \cite{DVEHK}.

\subsection{When there are no clocks in the complement}
\label{Subsection: N+0}

We have up to this point been assuming that $R^c$ is non-empty, i.e.\ that there are other clocks in the universe besides the ones that we are using. This is useful because it implies that the physical algebra $r(\mathcal{A}_{\mathcal{U}R}^H)$ is von Neumann on the physical Hilbert space. To see this, we can map this algebra into the perspective of a different clock $C_j\not\in R$, obtaining $\Pi_{|j}\mathcal{A}_{\mathcal{U}R}^H$ (see Eq.~\eqref{wrongQRF}). Since $\mathcal{A}_{\mathcal{U}R}^H$ is a von Neumann algebra acting on the kinematical Hilbert space {that commutes with $\Pi_{|j}$}, its projection onto $\mathcal{H}_{|j}$ is also a von Neumann algebra, and we may conclude that $r(\mathcal{A}_{\mathcal{U}R}^H)$ is a von Neumann algebra on the physical Hilbert space (with $r$  even being faithful \cite{DVEHK}).

When we are using all the clocks, $R=\{C_1,\dots,C_n\}$, there are no clocks in the complement available to make the above argument. In fact, in this case $r(\mathcal{A}_{\mathcal{U}R}^H)$ is \emph{not} a von Neumann algebra. To see this, note that the same algebra in the perspective of some clock $C_i$ is spanned by operators of the form $\Pi_{|i} a e^{-i(H_S+\sum_{j\ne i}H_j)t}
\Pi_{|i}$  with $a\in\mathcal{A}_{\mathcal{U}}\otimes \bigotimes_{j\ne i}\mathcal{B}(\mathcal{H}_j)$. Thus, its commutant on $\mathcal{H}_{|i}$ is generated by gauge-invariant operators of the form $\Pi_{|i} a'$, where $a'\in\mathcal{A}_{\mathcal{U}}'$. But the gauge-invariant subalgebra of $\mathcal{A}_{\mathcal{U}}'$ is just $\mathbb{C}\mathds{1}$ by the ergodicity of modular flow, so the commutant just consists of multiples of $\Pi_{|i}$. Mapping this back to the perspective-neutral level, one finds $r(\mathcal{A}_{\mathcal{U}R}^H)' = \mathbb{C}\mathds{1}$, and hence the bicommutant is just the full algebra of bounded operators on the physical Hilbert space:
\begin{equation}
    r(\mathcal{A}_{\mathcal{U}R}^H)'' = \mathcal{B}(\mathcal{H}_{\text{phys}}).
\end{equation}
Clearly $r(\mathcal{A}_{\mathcal{U}R}^H)''\ne r(\mathcal{A}_{\mathcal{U}R}^H)$, so $r(\mathcal{A}_{\mathcal{U}R}^H)$ is not a von Neumann algebra.

An operational interpretation of this result can be obtained from von Neumann's bicommutant theorem, which says that the bicommutant of an algebra is equivalent to its closure in some operator topology.\footnote{The weak, strong, ultraweak and ultrastrong topologies are all valid here.} This means that any operator in the bicommutant may be approximated to arbitrary precision by an operator in the original algebra. In the present case, this means that \emph{any} physical operator may be implemented to arbitrary precision by a combination of dressed observables and reorientations in $r(\mathcal{A}_{\mathcal{U}R}^H)$. There is a trace on $r(\mathcal{A}_{\mathcal{U}R}^H)''$ -- it is just the ordinary Hilbert space trace over $\mathcal{H}_{\text{phys}}$. The density operator of any pure physical state $\Ket{\phi}\in\mathcal{H}_{\text{phys}}$ is just $\Ket{\phi}\!\Bra{\phi}$, and its entropy is $0$.

CLPW referred to this as a `puzzle'~\cite{Chandrasekaran:2022cip}, saying: \emph{``the algebra of operators accessible to an observer in $P$ should not depend in this way on whether there is an observer in $P'$''} (here $P=\mathcal{U}$, $P'=\mathcal{U}'$). We do not necessarily share this sentiment. The constraint $H=0$ is a global one and thereby restricts what each observer has access to in a way that depends on all subsystems affected by the constraint. Indeed, the physical representation $r(\mathcal{A}_{\mathcal{U}R}^H)$, expressed in the perspective of $C_i\in R$, involves by Eq.~\eqref{Equation: reduced algebra} the projector $\Pi_{|i}$ in Eq.~\eqref{proj}, which encodes implicit information also about clocks outside of $R$. In fact, even in situations where some local structure in gravitational constraints is taken into account, it turns out that finding subregions that are independent of one another is rather hard \cite{Folkestad:2023cze,Soni:2024oim}.

In any case it is somewhat physically unrealistic to have access to all of the clocks in the universe, so we will continue in the rest of the paper to assume that $R^c$ is non-empty.

\section{Subsystem relativity and quantum frame transformations}
\label{Section: subsystem relativity}

As we have explained, to define a meaningful entanglement entropy for the subregion $\mathcal{U}$, one needs to describe it relative to one or more of the clocks. The picture one obtains clearly depends on the choice of clocks that we employ. This phenomenon is known as `subsystem relativity' \cite{Hoehn:2023ehz,AliAhmad:2021adn,delaHamette:2021oex,Castro-Ruiz:2021vnq}, and it has two manifestations.

As described above, for a given clock $C_i$, the physical, perspective-neutral Hilbert space $\mathcal{H}_{\text{phys}}$ is isometric via the reduction map $\mathcal{R}_i(\tau)$ with the reduced Hilbert space $\mathcal{H}_{|i}$:
\begin{equation}
    \mathcal{H}_{\text{phys}} \simeq \mathcal{H}_{|i} = \mathcal{R}_i(\tau)\mathcal{H}_{\text{phys}}.
\end{equation}
This is true for any $C_i$, which implies that the reduced Hilbert spaces for the various clocks are all isometric to one another:
\begin{equation}
    \mathcal{H}_{|i} \simeq \mathcal{H}_{|j} = V_{i\to j}(\tau_i,\tau_j)\mathcal{H}_{|i},
\end{equation}
where
\begin{equation}
    V_{i\to j}(\tau_i,\tau_j) := \mathcal{R}_j(\tau_j)\circ \mathcal{R}_i(\tau_i)^{-1}
\end{equation}
is the QRF transformation from $C_i$ to $C_j$ perspective. Explicitly, it is a controlled unitary \cite{Hoehn:2019fsy}:
\begin{equation}\label{QRFtransf}
     V_{i\to j}(\tau_i,\tau_j) = \int_\mathbb{R}\dd{t}\ket{t+\tau_i}_i\otimes\bra{\tau_j-t}_j\otimes e^{-i(H-H_i-H_j)t}
\end{equation}
This is an isometry for any values of $\tau_i,\tau_j\in\RR$, but note that for different such values we are mapping between reduced states when the clocks $C_i,C_j$ read different times.

Similarly, if we are using some fixed set of clocks $R$, then we can map the algebra $r(\mathcal{A}_{\mathcal{U}R}^H)$ into the perspective of any $C_i\in R$ using the isomorphism provided by the reduction map in Eq.~\eqref{Equation: reduced algebra}:
\begin{equation}
    r(\mathcal{A}_{\mathcal{U}R}^H) \simeq \mathcal{A}_{R_{\bar{i}}\mathcal{U}|i}^{\text{phys}} = \mathcal{R}_i(\tau) r(\mathcal{A}_{\mathcal{U}R}^H)\mathcal{R}_i(\tau)^\dagger,
\end{equation}
and the algebras in the perspectives of any two $C_i,C_j\in R$ are isomorphic via the QRF transformation:
\begin{equation}
    \mathcal{A}_{R_{\bar{i}}\mathcal{U}|i}^{\text{phys}} \simeq \mathcal{A}_{R_{\bar{j}}\mathcal{U}|j}^{\text{phys}} = V_{i\to j}(\tau_i,\tau_j)\mathcal{A}_{R_{\bar{i}}\mathcal{U}|i}^{\text{phys}}V_{i\to j}(\tau_i,\tau_j)^\dagger.
\end{equation}

The first version of subsystem relativity comes from keeping $R$ fixed, but going to the perspectives of different clocks $C_i,C_j\in R$. The isomorphisms just noted mean that this is a particularly weak form of subsystem relativity. Indeed, moving between the perspectives of different clocks $C_i,C_j\in R$ is just a change in description of the same physical subsystem. In line with this, the QRF transformation can be viewed as a unitary change of gauge. This does not mean, however, that the choice of QRF is equivalent to a choice of gauge; rather it is a convention of how to split kinematical degrees of freedom into redundant and non-redundant ones. 

As explained in \cite{Hoehn:2023ehz} (and further elaborated on in \cite{DVEHK}), an internal QRF perspective $\mathcal{R}_i(\tau)$ is nothing but a tensor product structure (TPS) (or direct sum thereof) on the physical Hilbert space $\mathcal{H}_{\rm phys}$, i.e.\ a \emph{physical} definition of a subsystem decomposition. The non-local QRF transformation~\eqref{QRFtransf} thus is a change of TPS on $\mathcal{H}_{\rm phys}$ and this means that different clock QRFs decompose the global system, as well as its subsystem $r(\mathcal{A}_{\mathcal{U}R}^H)$ above, in different ways into further subsystems. In typical quantum mechanical QRF setups, this naturally implies the QRF dependence of correlations and a range of thermodynamical properties, such as thermality and temperature \cite{Hoehn:2023ehz}. It will be interesting to extend these observations to a gravitational context and our exploration of gravitational entanglement entropies is a first step in this direction. 

The second version of subsystem relativity {is a much stronger manifestation of this QRF dependence of subsystem decompositions}, and concerns what happens when we change the set $R$ of clocks we are using. Clearly, the algebras for two different sets of clocks $R_1,R_2$ are distinct, which is simplest to see at the perspective-neutral level:
\begin{equation}
    r(\mathcal{A}_{\mathcal{U}R_1}^H) \ne r(\mathcal{A}_{\mathcal{U}R_2}^H).
\end{equation}
These algebras describe different sets of physical degrees of freedom. For example, in the case where $R_1$ is empty (i.e.\ we use no clocks), we have already shown that the left hand side above is trivial, $r(\mathcal{A}_{\mathcal{U}R_1}^H)=\mathbb{C}\mathds{1}$ if the modular flow is ergodic as for vacuum states. On the other hand, the right hand side is non-trivial for non-empty $R_2$. More generally, we have\footnote{Here, $\vee$ is the `join' of von Neumann algebras, defined such that $\mathcal{A}_1\vee\mathcal{A}_2$ is the smallest von Neumann algebra containing both $\mathcal{A}_1$ and $\mathcal{A}_2$. One can show that $\mathcal{A}_1\vee\mathcal{A}_2 = (\mathcal{A}_1'\cap\mathcal{A}_2')'$.}
\begin{align}
    r(\mathcal{A}_{\mathcal{U}(R_1\cup R_2)}^H) &= r(\mathcal{A}_{\mathcal{U}R_1}^H) \vee r(\mathcal{A}_{\mathcal{U}R_2}^H),\\
    r(\mathcal{A}_{\mathcal{U}(R_1\cap R_2)}^H) &= r(\mathcal{A}_{\mathcal{U}R_1}^H) \cap r(\mathcal{A}_{\mathcal{U}R_2}^H),
\end{align}
which captures the structure of the physical degrees of freedom which can be accessed using different sets of clocks. Observe in particular that if $R_1$ and $R_2$ have no overlap, then 
\begin{equation}\label{trivialoverlap}
    r(\mathcal{A}_{\mathcal{U}R_1}^H) \cap r(\mathcal{A}_{\mathcal{U}R_2}^H) = r(\mathcal{A}_{\mathcal{U}}^H) = \mathbb{C}\mathds{1}
\end{equation}
with the last equality only holding for ergodic modular flows as in de Sitter space. Thus, in that case, distinct sets of clocks do not share \emph{any} degrees of freedom whatsoever. {This is relevant because two observers in $\mathcal{U}$ with distinct clocks $C_1,C_2$ (cf.~Fig.~\ref{Figure: observers in subregion}) therefore describe the QFT degrees of freedom in their subregion by \emph{distinct} physical subsystems that overlap trivially. } However, note that there will in any state be fundamental correlations between the degrees of freedom measured by the two observers, since generally $[r(\mathcal{A}_{\mathcal{U}R_1}^H), r(\mathcal{A}_{\mathcal{U}R_2}^H)]\ne 0$ (because the field operators in one algebra will not commute with the field operators in the other). Hence, the two algebras do \emph{not} define {independent} subsystems.

\subsection{The frame dependence of gravitational entropy}

The two versions of subsystem relativity have two different sets of repercussions for the observer dependence of subsystem entropies.
Causality implies that observers travelling along wordlines in $\mathcal{U}$ (cf.~Fig.~\ref{Figure: observers in subregion}) will at most have access to the degrees of freedom in $\mathcal{U}$ and the analogue applies to the observers in the causal complement $\mathcal{U}'$. For the QFT (and graviton) degrees of freedom, this is essentially encapsulated by the timelike tube theorem \cite{Borchers1961,Araki1963AGO,Witten:2023qsv,Strohmaier:2023opz}, while for the clocks it depends on our operational assumptions because they are independent quantum mechanical degrees of freedom. 

Let us denote by $R_\mathcal{U}$ and $R_{\mathcal{U}'}$ all the clocks in $\mathcal{U}$ and $\mathcal{U}'$, respectively, and suppose for a moment that all observers in $\mathcal{U}$ somehow have access to all of $R_{\mathcal{U}}$, and similarly those in $\mathcal{U}'$ have access to all of $R_{\mathcal{U}'}$. The weaker version of subsystem relativity then entails that all observers in $\mathcal{U}$ and $\mathcal{U}'$ agree, up to non-local unitaries, on the description of their full regional physical algebras $r(\mathcal{A}_{\mathcal{U}R_{\mathcal{U}}}^H)$ and $r(\mathcal{A}_{\mathcal{U}'R_{\mathcal{U}'}}^H)$, respectively. This includes the full regional density operator, associated with some global physical state, and so the full regional entanglement entropy is then observer-independent. The various contributions to it coming from different subsystems will, however, typically look different relative to different frames owing to the non-locality of the QRF transformation~\eqref{QRFtransf}. If $\mathcal{U}$ is a region whose boundary is a horizon, such as a black hole interior or a static patch in de Sitter, this implies that horizon entropy is observer-independent.\footnote{{In semiclassical regimes, this horizon entropy can be expressed as a generalised entropy, where the area term comes from the relative entropy contribution in Eq.~\eqref{Equation: intro semiclassical entropy}, see \cite{Jensen:2023yxy} (and also \cite{Chandrasekaran:2022cip,Kudler-Flam:2023hkl,Kudler-Flam:2023qfl,DVEHK}) for the case of one observer in each of $\mathcal{U}$ and $\mathcal{U}'$. The observer-independence of the total horizon entropy means also that all observers will agree on the area contribution, provided the semiclassical regime holds in each of their perspectives (so that the area term can be defined).}}  Similarly, any two observers will agree on the entanglement entropy of any subset of QFT and clock degrees of freedom they jointly have access to, but the various contributions to it will typically appear differently.

However, a given observer Alice may not have access to all the clocks in her region $\mathcal{U}$, but only the one $C_{\text{Alice}}$ she carries along. Then she only has access to the algebra $r(\mathcal{A}_{\mathcal{U}C_{\text{Alice}}}^H)$ encoding the QFT relative to her clock frame and the entanglement entropy she will ascribe to $\mathcal{U}$ will be the one defined by the density operator in this algebra corresponding to a given global physical state. Even if she did have access to more clocks, this would be the entropy contribution she associates with just the QFT degrees of freedom in $\mathcal{U}$. A second observer Bob, travelling along a distinct worldline in $\mathcal{U}$ and carrying a clock $C_{\text{Bob}}\neq C_{\text{Alice}}$, will similarly assign the regional QFT entanglement entropy to the algebra $r(\mathcal{A}_{\mathcal{U}C_{\text{Bob}}}^H)$ he has access to. The strong version of subsystem relativity now implies that these observer algebras overlap trivially (for vacuum modular flows) and so the two observers see entirely different dressed QFT degrees of freedom. 

This brings us to
one of the main points of the paper: the QFT entanglement entropy of the subregion $\mathcal{U}$ depends on which set of clocks $R$ we are using, even when the state of the overall system is fixed. Since the QFT degrees of freedom include the gravitons, this means that gravitational entropy is QRF- and thus observer-dependent.
In a sense, this is not too surprising: different clocks will naturally have access to different sets of degrees of freedom, which carry different information. Our work makes this precise. 

It is also important to point out that this result is, in gravity, both essential (because of the triviality of the algebra when no clocks are used and the flow is ergodic), and entirely consistent with the general notion that all physical quantities should be defined in a relational manner, i.e.\ in relation to a given reference frame/observer. Indeed, as explained above, the dressed observables $\mathcal{O}_{C_i}^\tau(a)$ capture the behaviour of the operators $a$ from the perspective of the clock $C_i$. Here, we argue that the entanglement entropy is no different: it measures the information content of the region $\mathcal{U}$ from the perspective of a clock.

As an extreme example of this phenomenon, it can be the case that a semiclassical approximation is valid when dressing to one clock, while the antisemiclassical regime is valid when dressing to another, within a fixed physical state. Indeed, consider a state $\Ket{\phi}$ which from the perspective of a certain clock $C_1$, takes the form
\begin{equation}
    \ket*{\phi_{|1}} \approx \ket{\phi_S}\otimes \ket{\omega}_2 \otimes \ket{g}_{R^c}.
\end{equation}
Here $\ket{\omega}_2$ is a state for another clock $C_2$, while $\ket{g}_{R^c}$ is a state for the rest of the clocks $R^c = \{C_i\mid i \ne 1,2\}$. Let us pick $\ket{g}_{R^c}$ such that a semiclassical approximation holds when computing the entropy of $r(\mathcal{A}^H_{\mathcal{U}C_1})$. Thus, the relative entropy $S_{\text{rel}}(\phi_S||\psi_S)$ contributes to the entropy of $\mathcal{U}$ as measured by $C_1$. Let us simultaneously choose $\ket{\omega}_2$ to have energies contained within a very narrow window. Then, as explained in Sec.~\ref{sec_beyond}, the entropy of $r(\mathcal{A}^H_{\mathcal{U}C_2})$, i.e.\ the entropy of $\mathcal{U}$ as measured by $C_2$, contains \emph{no} contributions from the QFT degrees of freedom.

More generally, there can be many different sources for this entropy relativity. Another extreme example, which we will discuss in \cite{DVEHK}, arises when at least one of the observers uses a clock subject to a degenerate Hamiltonian whose degeneracy does not depend on the energy. This can lead to a superselection of $\mathcal{H}_{\rm phys}$ and $r(\mathcal{A}_{\mathcal{U}C_i}^H)$ across the degeneracy sectors. For example, doubly degenerate clock Hamiltonians arise typically in relativistic dispersion relations in the form $H_i = \pm p_i^2/2m_i$ and, depending on one's operational assumptions, this leads to a superselection across positive and negative frequency mode sectors \cite{Hoehn:2020epv}. This means that two observers who carry along two such relativistic clocks but only have access to their respective positive frequency sectors (the forward evolving branch), can only compare their descriptions in the overlap of the two corresponding superselection sectors, i.e.\ $\mathcal{H}_{C_i,+}\cap\mathcal{H}_{C_j,+}\subset\mathcal{H}_{\rm phys}$. The two observers can thus not compare their full respective density operators, but only `half' of each, and in general this means that the full QFT entanglement entropies accessible by the two clock frames will differ. Similar conclusions hold for any clocks that only exist on some subspace of the physical Hilbert space.

Entropy relativity will also typically arise for states in which the two clocks carried by the two observers in question are not isomorphic. For example, one observer could carry a monotonic clock $C_1$ (thus a translation group QRF), while the other carries a periodic clock $C_2$ (thus a $\rm{U}(1)$-QRF), such as a harmonic oscillator. The overall constraint will still be a translation group generator, which means that $C_2$ is an `incomplete' QRF; it has a reorientation isotropy group $H\simeq\ZZ$ and this means that it can only resolve properties of the QFT that are periodic themselves \cite{periodic,delaHamette:2021oex}. This leads to an additional averaging that will typically affect the entanglement entropy \cite{DVEHK}.

But clocks can also be non-isomorphic when they are both monotonic, e.g.\ when they feature different spectra $\sigma_i$. This can lead to differing degrees of fuzziness of their clock states~\eqref{clockstate} and hence a different `coarse graining' of the QFT degrees of freedom, which again leads to differences in the respectively observed entanglement entropies. For instance, Alice may carry an ideal clock, while Bob carries one with bounded spectrum. Depending on the presence of other observers, $r(\mathcal{A}_{\mathcal{U}C_{\rm Alice}}^H)$ may be of Type $\rm{II}_\infty$, while $r(\mathcal{A}_{\mathcal{U}C_{\rm Bob}}^H)$ is of Type $\rm{II}_1$, so that there will exist a maximal entropy state for the regional QFT for Bob but not for Alice. Entropy differences, however, also arise when the clock spectra are bounded in different ways.
This case will be illustrated in a simple example in the next section, along with the most general source of entropy relativity: differing entanglement structures between the perspectives owing to subsystem relativity. This source arises for \emph{any} two clocks, i.e.\ also for isomorphic ones.

\subsection{Gravitational interferometer: two  sources of entropy relativity}\label{sec_interferometer}

\begin{figure}
    \centering
    \begin{tikzpicture}[scale=0.5]
        \draw[fill=black!10] (-1,0) circle (3) node {\parbox{2cm}{\centering\large massive \\ object}};
        
        \draw[thick,blue,-{Stealth[scale=1.1,angle'=45]}] (6,-4) -- (6,-3)
            .. controls (6,-2) and (7,-2) .. (7,-1)
            -- (7,0.3);
        \draw[thick,blue] (7,0) -- (7,1)
            .. controls (7,2) and (6,2) .. (6,3)
            -- (6,4);
        
        \draw[thick,blue,-{Stealth[scale=1.1,angle'=45]}] (6,-4) -- (6,-3)
            .. controls (6,-2) and (5,-2) .. (5,-1)
            -- (5,0.3);
        \draw[thick,blue] (5,0) -- (5,1)
            .. controls (5,2) and (6,2) .. (6,3)
            -- (6,4);

        \draw[thick,blue,-{Stealth[scale=1.1,angle'=45]}] (13,-4) -- (13,0.3);
        \draw[thick,blue] (13,0) -- (13,4);

        \node[below] at (6,-4.3) {\large $C_2$};
        \node[below] at (13,-4.3) {\large $C_1$};
        \node[right] at (6,-3.5) {$\ket{0}_2$};
        \node[right] at (6,+3.5) {$\ket{\tau_2}_2+\ket{\tau_2+\Delta\tau}$};

        \draw [semithick,decorate,decoration={brace,amplitude=5pt,raise=-4,mirror}]
  (7.6,-1.7) -- (7.6,1.7);
        \node[right] at (7.6,0) {$\tau_2+\Delta\tau$};
        \draw [semithick,decorate,decoration={brace,amplitude=5pt,raise=-4}]
  (4.4,-1.7) -- (4.4,1.7);
        \node[left] at (4.4,0) {$\tau_2$};
    \end{tikzpicture}
    \caption{
        We consider a state where, in the perspective of a particular clock $C_1$, the state of a second clock $C_2$ is in a superposition of clock states. One can imagine starting with $C_2$'s time reading $0$, and then exposing it to a `gravitational interferometer', which splits the state of the clock into two branches; in one branch we bring the clock close to a massive object, while in the other we keep it far away (or we accelerate them by different amounts). Gravitational redshift causes the two branches to experience different amounts of time, i.e.\ Shapiro time delay, so after recombining the branches the final state is in the desired superposition. 
    }
    \label{Figure: interferometer}
\end{figure}
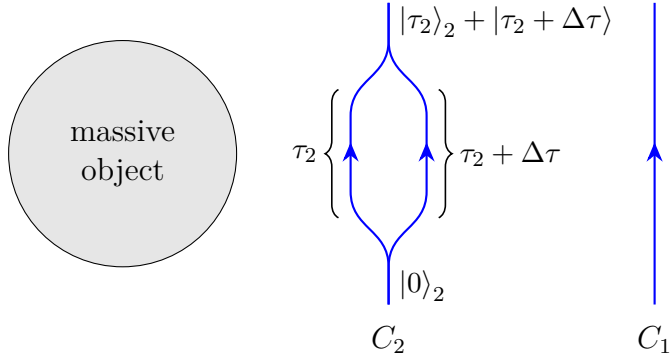

We will highlight two reasons why the QFT entropy of a physical state might differ between different observers: their clocks have different spectra, and the entanglement structure differs between the perspectives. The example under consideration will be such that in one clock's perspective, the wave function associated to another clock is in a superposition of two clock states. This can be interpreted as resulting from the use of a gravitational interferometer (see Fig.~\ref{Figure: interferometer}). In fact, we are looking at Shapiro time delay \cite{PhysRevLett.13.789}.

For this purpose, we consider the case of three non-ideal clocks with spectra bounded in both directions, hence the clock states~\eqref{clockstate} are normalisable \cite{Hoehn:2019fsy,DVEHK}, with perspectival states
\begin{align}
    \label{eq:state1}
    &\ket*{\phi _{\vert 1}} = \sqrt{N}\Pi_{|1} (\ket{\phi_S} \otimes \ket{f}_2 \otimes \ket{g}_3), \\
    &\ket*{\phi_{\vert 2}}  := V_{1 \to 2}(0,0) \ket*{\phi _{\vert 1}} = \sqrt{2\pi N}  f(-H_S-H_1-H_3) \Pi_{|2} (\ket*{\phi_S} \otimes \ket{0}_1 \otimes \ket{g}_3),
\end{align}
where $f(E)=\braket{E}{f}_2$ and $g(E)=\braket{E}{g}_3$  are normalised energy eigenbasis wave functions on their respective clock factors, while $N$ is a positive overall normalisation constant that fixes $\braket*{\phi_{|i}}{\phi_{|i}}=1$ for both $i=1,2$, and $V_{1\to2}$ is the QRF transformation~\eqref{QRFtransf}.
Both $\ket*{\phi _{\vert i}}$ represent the state when the clock $C_i$ reads 0.
We will additionally take $\phi_S$ to be canonically purified with respect to the KMS state $\psi_S$, which is also referred to as the former lying in the canonical cone of the latter state (e.g., see \cite[App.~C]{Jensen:2023yxy}). This is a technical restriction meaning that the modular conjugation operators $J_{\phi_S}$ and $J_{\psi_S}$, and likewise the relative modular conjugation operator $J_{\phi_S|\psi_S}$ are all the same. This is an insignificant restriction in the sense that any cyclic-separating state can be related to one in the canonical cone via a unitary in the commutant algebra. It also entails that $J_{\phi_S|\psi_S}H_SJ_{\phi_S|\psi_S} = -H_S$, which simplifies some expressions below.  We report more general expressions in the companion paper \cite{DVEHK} but here always make this `canonical cone' assumption for clarity.

One can show \cite{DVEHK} that the density operators for both states on their respective algebras $\mathcal{A}^\text{phys}_{\mathcal{U} \vert 1}$, $\mathcal{A}^\text{phys}_{\mathcal{U} \vert 2}$, which are both  type $\rm{II}_1$ factors, take the following form\footnote{To make contact with the generalised entropy computations done in  \cite{Chandrasekaran:2022cip} for a de Sitter static patch algebra, and in \cite{Jensen:2023yxy} for general bounded subregions, we might also consider the case where only two frames are present, with reduced state
$\ket*{\phi_{|2}}=N\Pi_2(\ket{\phi_S}\otimes \ket{f}_1)$ in the perspective of $C_2$, which, in contrast to the main text, would now represent a clock in $\mathcal{U}'$. Switching to the perspective of $C_1$ and computing the density matrix for $\mathcal{A}_{\mathcal{U}|1}^{\text{phys}}$ leads to
\[\rho_{\phi|1}:=Z_1 N f(\bar{H}_1)\Pi_{|1}\Delta^{1/2}_{\phi_S|\psi_S}
\Pi(\bar{H}_1,-\sigma_2)
\Delta^{1/2}_{\phi_S|\psi_S}\Pi_{|1}f^*(\bar{H}_1)e^{-\beta H_2}.
\]
If the complementary clock {$C_2$} is ideal, the projector in the middle can be ignored.  The QRF transformation~\eqref{QRFtransf} of this state recovers the one in \cite{Jensen:2023yxy} in the case of ideal clocks. Then under the semiclassicality assumptions of \cite{Chandrasekaran:2022cip,Jensen:2023yxy} the remaining operators all approximately commute, and the resulting entropy can be decomposed into a horizon area piece, plus entropy contributions associated with the observer and QFT fields in the region. Further details appear in \cite{DVEHK}, along with a discussion of the non-ideal clock case and how it generalises the corresponding ones in \cite{Chandrasekaran:2022cip,Jensen:2023yxy}. } 
\begin{align}
    \hspace{-5pt}\rho_{\phi \vert 1} &= Z_1 N e^{-\beta(H_2+H_3)} \Pi_{|1} \Delta^{1/2}_{\phi_S \vert \psi_S}
    \int_{\sigma_2}\dd \epsilon_2 \Pi(-\bar{H}_1-\epsilon_2,\sigma_3)
    |f(\epsilon_2)|^2|g(-\bar{H}_1-\epsilon_2)|^2  
    \Delta^{1/2}_{\phi_S \vert \psi_S}\Pi_{|1},\\
    \hspace{-5pt}\rho_{\phi \vert 2} &= Z_2 N e^{-\beta(H_1+H_3)} f(\bar{H}_2)
    \Pi_{|2}\Delta^{1/2}_{\phi_S \vert \psi_S}
    \int_{\sigma_1} \dd \epsilon_1 \Pi(-\bar{H}_2-\epsilon_1,\sigma_3)
    |g(-\bar{H}_2-\epsilon_1)|^2
    \Delta^{1/2}_{\phi_S \vert \psi_S}\Pi_{|2} f^*(\bar{H}_2),
\end{align}
where $Z_1,Z_2$ are defined in Eq.~\eqref{normal2}.
In these expressions, $\bar{H}_i := H_i - H$, and $\Pi(-\bar H_i-\epsilon_j,\sigma_3)$ is the projector restricting $-\bar H_i-\epsilon_j$ to eigenstates with eigenvalues in $\sigma_3$.
A major simplification occurs if we take $C_3$ to be effectively ideal, together with $g$ representing a normalised clock state of the form $N_3 \ket{\tau_3}_3$.  The former means that $\sigma_3$ is taken to be much larger (in both directions) than any other scale in the problem so that the projector onto the $\sigma_3$ range can be ignored, while the latter renders $|g|^2 =  \norm{\sigma_3}^{-1}$ completely constant, with $\norm{\sigma_i}:=\int_{\sigma_i}\dd{\epsilon}_i$ being the spectral range of $C_i$.  In this case, we obtain approximately
\begin{align}
    \label{eq:density1}
    \rho_{\phi \vert 1} &\approx  Z_1 N \norm{\sigma_3}^{-1} e^{-\beta(H_2+H_3)} \Pi_{|1} \Delta_{\phi_S \vert \psi_S}  \Pi_{|1},  \\
    \rho_{\phi \vert 2} &\approx Z_2 N\norm{\sigma_1} \norm{\sigma_3}^{-1} e^{-\beta(H_1 + H_3)} 
    f(-H_S-H_1-H_3) \Pi_{|2} \Delta_{\phi_S \vert \psi_S} \Pi_{|2} f^*(-H_S-H_1-H_3).
    \label{eq:density2}
\end{align}

At this stage some basic manifestations of subsystem relativity are already apparent:  the fact that \eqref{eq:density2} depends on the wavefunction $f$ while \eqref{eq:density1} does not, and the appearance of projector $\Pi_{|2}$ versus $\Pi_{|1}$ makes it obvious that the observable algebras $\mathcal{A}^{\text{phys}}_{\mathcal{U}|1}$ and $\mathcal{A}^{\text{phys}}_{\mathcal{U}|2}$ will differ in properties such as (but not limited to) their entanglement entropies, even in the same physical state. Indeed, by varying the wave function $f$, we can vary the entanglement entropy $S[\phi_{|2}]$ relative to clock $C_2$, while leaving the entanglement entropy $S[\phi_{|1}]$ relative to clock $C_1$ invariant. That this occurs is natural given that these are physically distinct (and, in fact, non-overlapping) algebras, cf.~Eq.~\eqref{trivialoverlap}.  But these algebras encapsulate all that can be observed about the QFT state on the subregion ``from the perspective'' of a given clock;  that these can differ dramatically based on both the clock state and the clock properties (here meaning the spectra), even when describing the same subregion, is physically significant.

Recalling that both algebras $\mathcal{A}^{\text{phys}}_{\mathcal{U}|i}$ here include not only the dressed QFT operators but also the reorientations of the clock $C_i$ (see Table \ref{Table: different algebras}), one can ask whether it is possible to further decompose these algebras' entropies into a simple sum of contributions associated with QFT and frames, separately. For states with arbitrary entanglement structure this is certainly not possible.  But states of the form \eqref{eq:state1} represent a case where such a decomposition might be expected, particularly when the projectors $\Pi_{|i}$ may be ignored, since then the frame and QFT factors are unentangled across unambiguous tensor factors \cite{DVEHK}.  Note, however, that in the case at hand, the algebras themselves do not respect this tensor product decomposition of the reduced Hilbert space (manifesting a property described in~\cite{WittenRevModPhys.90.045003}: entanglement is a property not just of states but of algebras).  The semiclassical conditions discussed in Sec.~\ref{sect:semiclassical} provide a scenario where such a decomposition is nevertheless possible.  In related works \cite{Chandrasekaran:2022cip,Jensen:2023yxy}, a semiclassicality condition was imposed on `slowly varying' frame wavefunctions (e.g.\ $f(\epsilon)$) that allowed the $\log$ of density matrices to be to approximately decomposed between the approximately commuting relative modular operator and wavefunction contributions.

Here we will instead consider a specific example state, which in $C_1$ perspective sees $C_2$ in a superposition of clock states.  Take as the wavefunction
\begin{equation}
    \ket{f}_2 = \sqrt{N_2} \qty\big(\ket{\tau_2}_2 + \ket{\tau_2 + \Delta \tau}_2), \quad f(E) = \sqrt{\frac{N_2}{2\pi}} e^{-i \tau_2 E} (1 + e^{-i \alpha E / \norm{\sigma_2}}),
\end{equation}
where $\alpha := \Delta \tau \norm{\sigma_2}$ is a dimensionless parameter. Note that $N_2$ depends on $\alpha$ because the clock states~\eqref{clockstate} are not orthogonal for non-ideal clocks. We can imagine the preparation of $\ket{f}_2$  by $C_2$ being sent through the two arms of an interferometer in a gravitational field with redshift or differing accelerations such that a time dilation by $\Delta\tau$ arises (see Fig.~\ref{Figure: interferometer}). This also implies that 
\begin{equation}
    \label{eq:stateV}
    \ket*{\phi_{\vert 2}} \propto \Pi_{|2} e^{iH_S \tau_2} \qty(\ket{\phi_S}\otimes\ket{-\tau_2}_1\otimes\ket*{\tau_3-\tau_2}_3 + e^{iH_S \Delta\tau}\ket{\phi_S}\otimes\ket{-\tau_2-\Delta\tau}_1\otimes\ket*{\tau_3-\tau_2-\Delta\tau}_3),
\end{equation}
which illustrates the general trade-off between superpositions (here in $\ket{f}_2$ relative to $C_1$) and entanglement (here in $C_2$ perspective) in QRF transformations \cite{Hoehn:2019fsy,Hoehn:2020epv,Giacomini:2017zju,Castro-Ruiz:2019nnl,AliAhmad:2021adn,Hoehn:2023ehz,delaHamette:2020dyi}.

Let us write down the expressions for the entropy. For the first density matrix \eqref{eq:density1}, we straightforwardly find the $f$-independent expression

\begin{equation}
    S[\phi_{\vert 1}] = - \log(Z_1 N \norm{\sigma_3}^{-1}) + \beta \expval*{H_2 + H_3}_{\phi_{\vert 1}} - \expval*{\log(\Pi_{|1} \Delta_{\phi_S \vert \psi_S}\Pi_{|1})}_{\phi_{\vert 1}}.
\end{equation}
For the second density matrix \eqref{eq:density2}, since $H_S$ and $\Delta_{\phi_S \vert \psi_S}$ do not commute, we will make use of the Baker-Campbell-Hausdorff-expansion by expanding in $\alpha \ll 1$; hence, the superimposed clock is close to a single clock state. This leads to\footnote{\label{footnote: tau dependence}Note that the $\tau_2$-dependence (not however the $\Delta\tau$-dependence) drops out by taking the expectation value.}
\begin{align}
    \label{eq:entropy2}
    S[\phi_{\vert 2}] = 
    &- \log(Z_2 N \norm{\sigma_1}\norm{\sigma_2}^{-1} \norm{\sigma_3}^{-1}) + \beta \expval*{H_1 + H_3}_{\phi_{\vert 2}}
    \nonumber \\
    &- \frac{1}{2} \left(\expval*{\log(\Pi_{|2}\Delta_{\phi_S(-\tau_2) \vert \psi_S}\Pi_{|2})}_{\phi_{\vert 2}} + \expval*{\log(\Pi_{|2}\Delta_{\phi_S(-\tau_2 - \Delta \tau) \vert \psi_S}\Pi_{|2})}_{\phi_{\vert 2}} \right) + \mathcal{O}(\alpha^2),
\end{align}
which through $\Delta\tau$ depends on $f$. At this point, we can discuss two sources of entropy relativity appearing in the QFT contributions to the full von Neumann entropies $S[\phi_{|1}]$ and $S[\phi_{|2}]$:
\begin{enumerate}
    \item $\alpha \to 0:$ in the limit when $f$ is also taken to be a single clock state, i.e.~$\Delta \tau \to 0$ with $\norm{\sigma_2}$ fixed,
    $\ket*{\phi_{\vert 2}}$ in \eqref{eq:stateV} is essentially of the same form as $\ket*{\phi_{\vert 1}}$ in \eqref{eq:state1} with $\phi_S$ additionally evolved forward in time. The $\tau_2$-dependence drops in the expectation value and in the end, the difference in entropy only stems from the different fuzzinesses of the clocks. Namely when $\sigma_1 \neq \sigma_2$, $\log(\Pi_{|1}\Delta_{\phi_S \vert \psi_S}\Pi_{|1})$ and $\log(\Pi_{|2}\Delta_{\phi_S  \vert \psi_S}\Pi_{|2})$ will functionally differ.\footnote{The two entropies can differ numerically based merely on the state-independent normalization constant(s) in the trace(s), which we have fixed in accordance with Section \ref{Subsection: normalisation and type}.  We refer here to a more substantial \textit{functional} difference in the entropies, based on the alternate course-grainings signified by $\Pi_{|1}$ and $\Pi_{|2}$, which lead to different sensitivities to changes in the QFT state.}
    \item $\sigma_1 = \sigma_2$: we can also look at the case when the two clocks are isomorphic such that the projected modular operators are functionally the same. Since $\alpha \neq 0$ we find that due to the different entanglement structure of $\ket*{\phi_{\vert 2}}$ in \eqref{eq:stateV} an extra term proportional to $i\alpha  \expval*{\comm*{\log (\Pi_{|2}\Delta_{\phi_S(-\tau_2) \vert \psi_S}\Pi_{|2})}{H_S}}_{\phi_{\vert 2}}$ appears if one Taylor expands the last term in \eqref{eq:entropy2}.
\end{enumerate}

So, the gravitational interferometer exemplifies two possible sources for the relativity of entropy: non-isomorphic clocks, and a different entanglement structure of the reduced states coming from the same global states $\Ket{\phi}$. This example will be further discussed in \cite{DVEHK}.

\section{Conclusion and outlook}
\label{Section: conclusion}

In this paper, we have explained how the formalism of quantum reference frames (QRFs) provides an appealing arena in which to understand recent developments involving observers in the regularisation of entropy in quantum gravity. We have argued that the observer invoked {by CLPW} in~\cite{Chandrasekaran:2022cip} (and by related works \cite{Jensen:2023yxy,Kudler-Flam:2023hkl,Kudler-Flam:2023qfl,Witten:2023qsv,Witten:2023xze}) is indeed a QRF, as exemplified by the slogan PW = CLPW (see Eq.~\eqref{Equation: PW=CLPW}) {The left hand side refers to the Page-Wootters formalism \cite{Page:1983uc,1984IJTP...23..701W}, which (in suitably generalised form) is a key ingredient of the \emph{perspective-neutral} approach to QRF covariance \cite{delaHamette:2021oex,Hoehn:2023ehz,Hoehn:2019fsy,Hoehn:2020epv,AliAhmad:2021adn,Giacomini:2021gei,Castro-Ruiz:2019nnl,delaHamette:2021piz,Vanrietvelde:2018dit,Vanrietvelde:2018pgb,Hoehn:2021flk,Suleymanov:2023wio}. W}e have used this insight to generalise and expand upon those previous results. Hopefully the reader has been convinced that it is both natural and fruitful to consider the role played by QRFs in quantum gravity.

A particular focus of our work was a system with an effective QFT coupled to multiple observers/QRFs in the $G_N\to0$ limit, each carrying its own clock, and we studied the way in which the gravitational entropy of a subregion can depend on which observers are being used. We noted that the total entropy of \emph{all} the degrees of freedom in a subregion, including all the observers, does not depend on which observer is used to compute it -- because each observer sees the same algebra (up to the unitary QRF transformations we described in Sec.~\ref{Section: subsystem relativity}). Thus, horizon entropies such as the black hole entropy~\cite{Kudler-Flam:2023hkl} are observer-independent. On the other hand, if we are interested in the experience of a particular observer (or subset of observers) in the subregion, operational considerations mean one should restrict to the degrees of freedom accessible to these observers alone -- and, due to subsystem relativity, it is the entropy of this restricted set of degrees of freedom which depends on the choice of observers. {In particular, what two distinct observers, equipped with distinct clocks and travelling along different worldlines within the same subregion (cf.~Fig.~\ref{Figure: observers in subregion}), `see' as the entanglement entropy of the QFT degrees of freedom (including gravitons) is highly observer-dependent and we have provided examples of this. This is what we mean by the observer dependence of gravitational entropy. }

We have tried to provide a relatively approachable account of our results, but in doing so we have had to skip over many important details. We refer the interested reader to our longer companion paper~\cite{DVEHK}, in which we provide a much more comprehensive account of the QRF-dependence of gravitational entropy, including mathematical subtleties which we have only hinted at here, as well as a few more explicit examples of the phenomenon.

Before ending the paper, let us list a few possible future directions.

First, it would be interesting to understand how the thermodynamics of a gravitational subregion might depend on the choice of observers, which is certainly the case in simpler quantum mechanical systems~\cite{Hoehn:2023ehz}. Horizon entropies are known to obey a generalised second law~\cite{Wall_2012}, a proof of which has recently been extended to the current setup of von Neumann algebras and observers \cite{Faulkner:2024gst,Ali:2024jkx,KirklinGSL}. But, as we stated above, horizon entropies are not observer-dependent. It is thus natural to ask whether a generalised second law can be proven for the entropy of the subsystem consisting of the degrees of freedom in a subregion relative to some \emph{subset} of observers, instead of the full horizon entropy. One complicating factor here is that, in the presence of interactions between the observers and the fields, such a subsystem can freely exchange heat with its complement (unlike in the case of the horizon entropy, where causality implies heat can only radiate \emph{out} of the subsystem). Presumably, however, since we are neglecting any such interactions, a version of the generalised second law can be proved {also for subsets of observers}.

Of course, observers do interact with the fields. Indeed, to a certain extent it is these interactions which allow the observers to measure the fields at all: each observer carries some measurement apparatus whose state evolves according to its coupling with the QFT (this is the subject of quantum measurement theory~\cite{QuantumMeasurement,fewster2023measurement,Fewster:2018qbm}). It would be interesting to see what happens to our results once one turns on these interactions by changing the constraint via $H\to H+H_{\text{int}}$. The physical Hilbert space of a gauge theory can change drastically under an arbitrarily small perturbation to the constraints {(e.g., see \cite{Marolf:2009wp} for a simple example)}, so understanding this could be quite complicated. {First steps to study interacting quantum reference frames have been taken in \cite{Hohn:2011us,Smith:2019imm,Smith2019quantizingtime,Castro-Ruiz:2019nnl,Hoehn:2023axh}}. 

Another feature of gravity that we have ignored is that its gauge group is vastly more complicated than $\RR$. Indeed, the full gauge group consists of all spacetime diffeomorphisms, and more general QRFs than those we have studied here will transform under more than just the single boost diffeomorphism we have invoked. {(A treatise of classical dynamical frames in gravity can be found in \cite{Goeller:2022rsx,Carrozza:2022xut} and some extension to the quantum realm appears, e.g., in \cite{Kabel:2023jve,Kabel:2024lzr,Blommaert:2019hjr,Nitti:2024iyj}.)} Physically speaking, these QRFs can carry not just clocks, but also rulers, and tetrads, and so on. This would then be another source of observer dependence in the entropy: one observer could carry just a clock and a ruler, another could carry a clock and a tetrad, etc. Depending on what the observer carries, it will be able to witness different field observables, and thus measure a different entropy. By considering QRFs transforming under more general groups than just $\RR$~\cite{delaHamette:2021oex}, we hope to address this in forthcoming work (see also~\cite{Fewster:2024pur}). 

We anticipate that subsystem relativity should generalise in some form to QRFs associated with the diffeomorphism group and thereby cause a generic entropy relativity in gravity. This is presumably connected with the recent observation that events and localisation in quantum gravity are observer/QRF-dependent \cite{Kabel:2024lzr,Nitti:2024iyj}, as this means that different QRFs will not agree on what the subsystem is.

Finally, although we have sometimes used in this paper a heuristic picture of QRFs evolving along worldlines, a QRF need not be of this nature. In particular, in gravity and gauge theory it is very natural to use \emph{dynamical edge modes} as reference frames~\cite{Carrozza:2021gju,Carrozza:2022xut,Goeller:2022rsx,Araujo-Regado:2024dpr,Kabel:2023jve,Gomes:2016mwl}. These are degrees of freedom which do not live on a worldline in the subregion, but rather on the \emph{boundary} of the subregion. The similarity between the construction of the extended phase space used in the study of classical edge modes~\cite{Donnelly_2016}, and the crossed product von Neumann algebra described in~\cite{Chandrasekaran:2022cip} and the present work, has already been pointed out~\cite{Jensen:2023yxy,Klinger:2023tgi,Klinger:2023auu,Ciambelli:2023mir,Elliot}. There are different sets of dynamical edge modes one could use on any given boundary \cite{Carrozza:2022xut,Carrozza:2021gju,Araujo-Regado:2024dpr}. Much as a choice of clock in this paper yields a particular physical representation of the crossed algebra, a choice of dynamical edge modes yields a particular map from the physical phase space of a classical field theory to its extended phase space. It would be interesting to understand this from the point of view of QRFs, and in light of the results of this paper. Indeed, in the quantum theory different choices of dynamical edge modes are different choices of QRF, and so should result in different entropies.

\section*{Acknowledgements}

\noindent We thank Gon\c{c}alo Ara\'ujo Regado, Wissam Chemissany, \r{A}smund Folkestad, Laurent Freidel, Elliot Gesteau, Ted Jacobson, Fabio Mele, Federico Piazza, Gautam Satishchandran and Antony Speranza for helpful discussions and comments.
PH is grateful for the hospitality of the high-energy physics group at EPFL Lausanne, where part of this work was carried out. This work was supported by funding from Okinawa Institute of Science and Technology Graduate University. This project/publication was also made possible through the support of the ID\# 62312 grant from the John Templeton Foundation, as part of the \href{https://www.templeton.org/grant/the-quantum-information-structure-of-spacetime-qiss-second-phase}{\textit{`The Quantum Information Structure of Spacetime'} Project (QISS)}.~The opinions expressed in this project/publication are those of the author(s) and do not necessarily reflect the views of the John Templeton Foundation. Research at Perimeter Institute is supported in part by the Government of Canada through the Department of Innovation, Science and Economic Development and by the Province of Ontario through the Ministry of Colleges and Universities.

\printbibliography

\end{document}